%% file: mainpaper.tex
\PassOptionsToPackage{table}{xcolor}
\documentclass[VANCOUVER,STIX1COL,table]{WileyNJD-v2}

\articletype{Research Paper}

\usepackage[table]{xcolor}
\usepackage{multirow}
\usepackage{multicol}
\usepackage{tabularx}
\usepackage{listings}
\usepackage{graphicx}
\usepackage{soul}
\usepackage{csquotes}
\usepackage{amssymb}
\usepackage{amsmath}
\usepackage{paralist}
\usepackage{fixltx2e}
\usepackage{fancyvrb}
\usepackage[final]{fixme}
\usepackage{siunitx}
\usepackage{url}
\usepackage{caption}
\usepackage{pifont}
\usepackage{cleveref}
\usepackage{booktabs}
\usepackage{adjustbox}
\usepackage{xspace}
\usepackage{subfig}

\fxsetup{footnote,nomargin}
\definecolor{lightgray}{gray}{0.9}

\soulregister\cite7
\soulregister\ref7
\soulregister\pageref7

\Crefname{hyp}{Hypothesis}{Hypotheses}

\newtheorem{obs}{Observation}
\Crefname{obs}{Observation}{Observations}

\graphicspath{{figures/}}

\newcommand{\daikon}{\textsc{D}aikon\xspace}
\newcommand{\techname}{\textsc{IPA}\xspace}

\newcommand{\spara}[1]{\vspace*{0.05in}\noindent\textbf{#1}}

\begin{document}
%
\title{Error Propagation Analysis for Multithreaded Programs: An Empirical Approach}


\author[1]{Stefan Winter*}
\author[2]{Abraham Chan}
\author[3]{Habib Saissi}
\author[2]{Karthik Pattabiraman}
\author[4]{Neeraj Suri}

\authormark{STEFAN WINTER \textsc{et al}}

\address[1]{\orgdiv{SoSy-Lab}, \orgname{LMU Munich}, \orgaddress{\country{Germany}}}
\address[2]{\orgdiv{Department of Electrical and Computer Engineering}, \orgname{University of British Columbia}, \orgaddress{\state{British Columbia}, \country{Canada}}}
\address[3]{\orgname{AUSY Technologies}, \orgaddress{\state{Frankfurt}, \country{Germany}}}
\address[4]{\orgdiv{School of Computing and Communications}, \orgname{Lancaster University}, \orgaddress{\country{UK}}}

\corres{*Stefan Winter. \email{sw@stefan-winter.net}}

\input{abstract}

\keywords{Error Propagation Analysis, Fault Injection, Concurrency, Multithreading}

\maketitle

\input{intro}
\input{background}

\input{methodology}

\input{evaluation}
\input{discussion}
\input{conclusion}
\input{acknowledgements}

\input{mainpaper.bbl}
\end{document}

%% file: abstract.tex
\abstract{
Fault injection is a technique to measure the robustness of a program to errors by introducing faults into the program under test.
Following a fault injection experiment, Error Propagation Analysis (EPA) is deployed to understand how errors affect a program's execution.
EPA typically compares the traces of a fault-free (\emph{golden}) run with those from a faulty run of the program.
While this suffices for deterministic programs, EPA approaches are unsound for multithreaded programs with non-deterministic golden runs.
In this paper, we propose Invariant Propagation Analysis (IPA) as the use of automatically inferred \emph{likely invariants} (``invariants'' in the following) in lieu of golden traces for conducting EPA in multithreaded programs.
We evaluate the stability and fault coverage of invariants derived by IPA through fault injection experiments across six different fault types and six representative programs that can be executed with varying numbers of threads.
We find that stable invariants can be inferred in all cases, but their fault coverage depends on the application and the fault type.
We also find that fault coverage for multithreaded executions with IPA can be even \emph{higher} than for traditional singlethreaded EPA, which emphasizes that IPA results cannot be trivially extrapolated from traditional EPA results.}


%% file: intro.tex
\section{Introduction}

Software fault injection (SFI) \cite{Duraes2006,Natella2013,Aliabadi2014} is a widely-used technique for testing the robustness of software to faults in its operational environment.
For this purpose, SFI introduces faults into the software under test (SUT) and its environment, executes the SUT, and observes its behavior under the faults.
While the nature of SFI closely resembles that of mutation analysis, the faults considered in SFI are not limited to simple syntactical mutation operators, but also include more complex emulations of real world faults (e.g., software bugs)\footnote{Contrary to mutation testing, there is no established \emph{coupling hypothesis} \cite{DeMillo1978.coupling,Offutt1989.coupling,Offutt1992.coupling} in SFI.}.
Similar to other types of robustness tests, such as fuzzing approaches, SFI relies on \emph{negative oracles} to determine if a test passes.
Unlike traditional tests that pass if the SUT's output matches an expected output, SFI tests pass if certain undesired behaviour does \emph{not} occur (e.g., program crashes).

An important type of negative oracles is \emph{error propagation}, i.e., the corruption of a module's internal state by an injected fault.
In general, error propagation is undesirable because it can lead to failures that are hard to diagnose and recover from.
The identification of how such state corruptions evolve from a fault activation in execution is referred to as \emph{error propagation analysis} (EPA).
EPA typically requires capturing a detailed execution trace of the program when running a test.
After the termination of a test, its execution trace is compared to an execution trace from a fault-free execution, also known as the \emph{golden run} \cite{Christmansson1998,Hiller2002,Leeke2009}).
Any deviation from the golden run is considered to be an instance of error propagation.
While other forms of EPA exist, in this paper, we refer to golden-run based EPA when we say EPA as it's the dominant form.

While the golden run comparison technique works well for EPA of deterministic programs and execution environments, it can lead to spurious outcomes in the presence of non-determinism, which can cause trace deviations that do \emph{not} indicate error propagation. 
One of the primary sources of non-determinism is multithreading in programs, which is becoming more prevalent as processors attain increasing core counts.
In general, there are two cases of non-determinism that occur in multithreaded programs: \emph{scheduling} and \emph{data non-determinism}.
Since threads can execute in different orders due to non-deterministic scheduling, the traced values in the program may differ across program executions.
However, the value of a variable within a thread does not change -- this is called scheduling non-determinism. 
On the other hand, data non-determinism (i.e., a race condition) occurs when multiple threads write to a shared variable in different orders,
and hence the variable's value differs from one execution of the program to another. 
Scheduling non-determinism may be addressed by tracking the thread identifiers and separating the traced values based on the identifiers.
However, data non-determinism cannot be mitigated by tracking thread identifiers as the variable's value is a function of the specific interleavings among threads. 
In practice, these are a large number of possible thread interleavings and it is not practical to track the values in each interleaving.
Thus, an alternative EPA framework is required for handling data non-determinism in shared memory multithreaded programs, which is the focus of this paper. 
 
To mitigate the effects of spurious EPA results due to multithreaded executions, we propose the use of dynamically generated \enquote{likely} invariants \cite{Ernst2000} to perform EPA for multithreaded programs.
An invariant is a property on a program's data values that holds across \emph{all} of its executions.
\emph{Likely invariants} are those that hold across some executions of the program but do not necessarily hold across others.
Ernst et. al.~\cite{Ernst2000} proposed the idea of likely invariants that are automatically derived from execution traces of a program, and used it for program comprehension. Since the publication of this seminal work, likely invariants have been used for many dependability-related tasks, such as error detection~\cite{Schuler2009}, fault localization~\cite{Sahoo2013}, program repair~\cite{Perkins2009} and test case generation~\cite{Xie2006}.
\emph{However, to the best of our knowledge, there has been no systematic study on the use of dynamically generated likely invariants for EPA.}

There are three reasons why likely invariants are a good fit for the EPA problem.
First, likely invariants can be automatically generated by analyzing the traces from different executions of a program, without any programmer intervention.
This is critical for the technique to scale to large, real-world applications.
Second, likely invariants are often compact, and can be checked with low overhead at run-time, e.g., as predicates for executable assertions.
This makes them easily applicable as oracles.
Thirdly, and most importantly, likely invariants can be conditioned such that they hold across the entire set of executions on which the program is trained, automatically abstracting out the non-deterministic parts of the program. 
This makes them especially well-suited for handling data non-determinism in performing EPA of multithreaded programs. 

However, likely invariants characterize correct executions with less precision than true invariants~\cite{Ernst2000}, which may reduce their efficacy for EPA.
Consequently, the question we ask in this paper is: \enquote{\emph{How effective are the invariants\footnote{From this point on, when we say invariants, we mean likely invariants.} generated by automated techniques in tracking error propagation in multithreaded programs?}}.
It is important to answer this question to determine if existing invariant generation techniques are sufficient for EPA, or if new techniques need to be developed.
We experimentally measure the effectiveness of an invariant in terms of two attributes,
\begin{inparaenum}[(1)]
\item the \emph{stability} of the generated invariant set across different (non-deterministic) executions of the program, and
\item the \emph{fault coverage} of the generated invariants for different fault types, corresponding to common software faults.
\end{inparaenum}

We make the following contributions in this paper:
\begin{itemize}
\item We propose the use of invariants for performing EPA in multithreaded programs.
\item We build a framework called \techname (Invariant-based Propagation Analysis) to derive dynamic invariants for multithreaded programs through an automated, end-to-end process (Section~\ref{sec:soba}).
\item We empirically assess the efficacy of the invariants derived using \techname for six representative multithreaded programs through fault-injection experiments (Section~\ref{sec:evaluation}).
\end{itemize}

In the conference version of this paper~\cite{Chan2017}, we presented \techname and an experimental evaluation of the effectiveness of likely invariants for detecting faults in multithreaded programs with fault injection experiments.
In this paper, we present an expanded evaluation (Sections~\ref{sec:confidence} \& \ref{sec:granularity}), with two previously unexplored parameters in our \techname framework - the confidence and granularity of the dynamicly inferred invariants, a side-by-side comparison of the fault injection results of the benchmark applications executed with a single thread versus multiple threads, a formal system model of EPA to demonstrate why golden run based EPA is unsound in multithreaded programs, and a fuller justification for the fault model used.

Our results are as follows.
We find that the traditional form of EPA is unsuitable for multithreaded programs due to their non-determinism.
We also find that the invariants derived by IPA are stable across multiple executions, and provide coverage ranging from 10\% to 97\% depending on the fault type and program.
Then, we find that the proposed IPA framework is substantially faster than traditional EPA-based analysis, while incurring a 2-90\% one time setup overhead.
Finally, we perturb the confidence and granularity parameters of \techname to determine whether the fault coverage can be improved.
We find no significant difference in the number of inferred invariants when confidence is lowered and that invariants inferred at the basic block granularity do not stabilize within 20 runs, making them unsound for EPA.
Our results thus indicate that invariants offer a promising alternative to EPA in some programs but not others, and the results are highly dependent on the program.


%% file: background.tex
\section{Background and Related Work}
\label{sec:background}

In this section, we first describe the notions of fault injection and EPA.
Then, we show an example of why golden run EPA is unsound for multithreaded programs.
Finally, we describe likely invariants and survey related work in the field on using likely invariants.

\subsection{Fault Injection} 
\label{sec:fault-injection}

Software fault injection is a technique to emulate bugs by modifying one or more components of a software system.
It has been widely deployed to advance test coverage and software robustness by exploring error handling paths of programs (e.g., \cite{Koopman2000,Fetzer2004,Fu2005,Fu2007,Marinescu2009}).
There are two categories of fault injection: compile-time injections and run-time injections.
Compile-time injections typically involve modifying source code (e.g., SAFE \cite{Natella2013}) or binary code (e.g., G-SWFIT \cite{Duraes2006} or EDFI \cite{Giuffrida2013}), similar to mutation testing.
In contrast, run-time injections mimic software events that corrupt instructions and memory at run-time.
The sensitivity of programs to such events is difficult to assess through traditional testing techniques~\cite{Sullivan1991}.
We focus on run-time injections, and refer to these as fault injections in this paper.

Traditionally, fault injection tools have targeted hardware faults, such as single event upsets caused by particle strikes on chips.
However, an increasing number of fault injection systems now target software faults.
Fault injection systems, such as FIAT \cite{Segall1988}, LLFI \cite{LLFI}, or PDSFIS \cite{Jin2008},
explicitly support the emulation of a wide range of software faults at run-time.
For instance, buffer overflow errors can be simulated by under-allocating malloc calls by some number of bytes.
Other examples include simulating invalid pointer errors by randomly corrupting pointer addresses, and race conditions by acquiring non-existent or incorrect locks. 
Such simulated software bugs can be injected at either random or specific program points in order to study their effects on a program. 

\subsection{Error Propagation Analysis (EPA)} 
\label{sec:error-prop-analys}

The effects of a software fault depend on both its type and the location in which it occurs. 
EPA attempts to answer the following question:
\emph{``How does an injected fault of known type and known location propagate within a program?''}

Existing EPA approaches in tools such as PROPANE~\cite{Hiller2002} or LLFI~\cite{LLFI} make use of either instruction or variable trace comparisons between golden and faulty runs of programs.
Deviations between traces can be interpreted as data violations or control flow violations.
Data violations occur when identical instructions at the same program point are invoked with different values.
Control flow violations occur when the instruction orders differ. 
Either violation is considered an indication of a software fault.
However, this approach assumes that traces from golden runs are identical as long as the program is operating on the same inputs.
Any non-determinism in the program can violate this assumption, such as that caused by multithreading.

Lemos et. al.~\cite{Lemos} addressed the non-determinism problem in EPA using approximate comparison techniques used in computational biology (e.g., DNA sequencing) to compare golden traces and faulty traces.
This approach, however, does not compare the non-deterministic portions of the trace with the golden run, effectively limiting its coverage.
Unfortunately, the traces of multithreaded programs tend to be non-deterministic for the largest part.

Leeke et al.~\cite{Leeke2009} attempt to solve the non-determinism problem in EPA using a \emph{reference model}, which is a statistical characterization of the system's outputs.
At a  high-level, reference models are similar to likely invariants.
However, unlike likely invariants, which can be automatically derived, the reference model requires significant manual effort and also detailed domain knowledge.
Further, for many systems, it may not be possible to derive a reference model if the outputs do not conform to well-known statistical distributions.

TraceSanitizer \cite{tsan} is another recent technique\footnote{Most of the authors of this paper overlap with the TraceSanitizer paper.} to solve the problem of spurious deviations in EPA execution traces resulting from multithreading and dynamic memory allocations.
Unlike IPA, the comparison of traces processed by TraceSanitizer can be proven to provide sound EPA for certain types of programs.
While IPA cannot provide such guarantees, it does not suffer from TraceSanitizer's applicability constraints, and can hence be used for any type of program.
Moreover, since IPA does not rely on SMT solving like TraceSanitizer does, 
it better scales for programs with highly complex inputs, for which TraceSanitizer's SMT formula construction time can become large.
Further, it does not suffer from the unpredictable and potentially large SMT solving times observed in Trace Sanitizer \cite{tsan}.

A number of fault injection tools for large and complex systems face similar problems of non-determinism as the ones that motivated our work.
FATE \cite{Gunawi2011} injects combinations of errors at the interface between service providing software and its interface to system resources, such as I/O devices, to assess recovery mechanisms. 
However, it may falsely conclude recovery failures when the timing between injections and recovery checks is insufficient.
To address this issue, FATE relies on manual specifications using a dedicated specification language, which requires programmer effort.
In contrast, our approach is completely automated. Further, FATE does not investigate how errors propagate like we do in this paper.

Deligiannis et al. \cite{Deligiannis2016} report on non-determinism as a challenge for distributed system testing and propose a systematic approach 
to explore the various behaviors that can result from such non-determinism during test.
Cotroneo et al. \cite{Cotroneo2020} account for non-deterministic event traces in fault injection campaigns for distributed systems by building anomaly detectors 
based on fault-free executions based on Markov chains that model event orders.
They solve a similar problem as us, but the conceptual difference between the execution traces renders a direct adoption of their approach for our domain (and vice versa) infeasible.
This is because in distributed systems, the (binary) occurrence and order of events is of high relevance, but modeling all possible events of highly parameterized operations as in program statements would lead 
to a prohibitively large alphabet and modeling each program statement as an event to prohibitively complex models.
Similarly, data invariants as in our work are oblivious of protocol-level errors in the communication between nodes in a distributed system, and hence cannot aid in their detection.

\subsection{EPA in Multithreaded Programs}
\label{sec:motivating-example}

In this section, we provide an example of why EPA using golden run comparisons does not suffice for multithreaded programs as a result of \emph{data non-determinism}.

Consider the function in Figure~\ref{fig:func_example}, created for the purpose of this example.
The function takes a single input $x$ and adds it to a work queue \texttt{workChunks}.
The elements in the work queue are processed by separate threads and there is no order dependency across any two elements, i.e., the order in which elements are added to or taken from the queue is not functionally relevant for the program.
If 4 work chunks \texttt{0,1,2,3} were to be added to \texttt{workChunks}, a golden run trace of a single threaded run would invoke \texttt{addChunk} with these four values in sequence.
As long as the work chunks do not change, such a golden run trace is consistent for every execution of the program.

Consider the case where the program is executed on four threads for the same set of chunks.
Because the thread schedule can change, the order in which chunks are added to the queue can change as well, resulting in different execution traces for the program and different heap memory states for \texttt{workChunks}.
For example, the 4 work chunks could be added in reverse order, leading to \texttt{workChunks} containing \texttt{[3,2,1,0]} instead of \texttt{[0,1,2,3]} in the single-threaded execution discussed before.
These deviations would erroneously trigger an alarm by classical golden-run based EPA and cannot be resolved by simply tracking thread IDs, because the thread scheduling deviations lead to deviating \emph{memory states}.

The approach presented in this paper circumvents this problem in two ways.
First, it does not rely on a comparison of execution orders.
Therefore, different instruction orders resulting from different thread schedules cannot affect a comparison between fault-free and faulty runs.
Second, it does not rely on unconditional data comparisons.
Therefore, even if variable values, such as that of \texttt{workChunks}, differ across executions, the invariants that our approach learns for the comparisons across executions are \emph{conditional}, 
and reflect the changes in the data structures.
For instance, when \texttt{addChunk(0)} is invoked, the approach is capable of learning that after the execution of that function, the element \texttt{0} is in \texttt{workChunks}, irrespective of what other elements it may contain and in which order.
We will detail this invariant-based approach in the following section.

\begin{figure}\
\begin{footnotesize}
\begin{Verbatim}[frame=single,fontsize={\scriptsize},numbers=left,numbersep=5pt,xleftmargin=10pt]
sem_t freeSlots; // initialized with SIZE
int nextIndex = 0;
int workChunks[SIZE];
pthread_mutex_t mutex;

int addChunk (int x) {
    sem_wait(&freeSlots);        
    pthread_mutex_lock(&mutex);  
    workChunks[nextIndex] = x;  // chunk order irrelevant
    nextIndex += 1;
    pthread_mutex_unlock(&mutex);
}
\end{Verbatim}
\end{footnotesize}
\caption{Example thread-safe function working on shared global data.}
\label{fig:func_example}
\end{figure}

\subsection{Likely Invariants}
\label{sec:likely-invariants}

\emph{True invariants} are predicates that are valid across the set of all executions of a program. 
Therefore, the violation of a true invariant necessarily indicates the presence of a fault,  provided the invariant was inferred from a correct program.
Thus, true invariants are sound, but not necessarily complete indicators for error propagation.
Unfortunately, the existence of such true invariants is undecidable in the general case \cite{Padon2016}, 
which makes their automated inference difficult, if not impossible.

\emph{Likely invariants}, in contrast, only hold for observed executions but not necessarily for all executions.
Thus, they may contain spurious invariants in addition to true invariants. 
Further, likely invariants may not comprise all true invariants as some true invariants may not be exercised in the set of observed executions. 
Consequently, likely invariants are both incomplete and unsound in the general case, and hence incur both false negatives \emph{and} false positives.

Although likely invariants, unlike true invariants, bear a risk of false positives, we assert that this risk is substantially lower than for golden run comparisons in non-deterministic programs.
This is because EPA is typically done over a set of known inputs, and we only require that the likely invariants are stable over this set.
Further, likely invariants can be generated through automated techniques~\cite{Ernst2000, DIDUCE, DySy}, which make them a viable option even for highly complex programs.

In this paper, we focus on the likely invariants generated by \daikon~\cite{Ernst2000}, which is the most widely used likely invariant inference engine.
\daikon infers and reports likely invariants based on a set of execution traces.
DySy~\cite{DySy} and DIDUCE~\cite{DIDUCE} are other examples of dynamic invariant generation tools.
DySy first applies symbolic execution and then observes dynamic execution traces to generate invariants.
DIDUCE detects invariants and subsequently checks their violations to help programmers locate bugs.
However, all three systems suffer from the effects of thread non-determinism~\cite{Kusano2015}, rendering them unsuitable for multithreaded programs.
In recent work, Kusano et al.~\cite{Kusano2015} addressed this problem by developing a custom interleaving explorer for multithreaded programs called \enquote{Udon}.
However, Udon is not used for the purpose of EPA, which is our focus.
Our framework builds on top of Udon for invariant inference.

Prior work has used likely invariants for mutation testing and error detection.
For example, Schuler et al.~\cite{Schuler2009} assess the viability of invariant checking in mutation testing.
They find that an invariant approach yields a 97\% detection rate in their mutation experiments.
However, they evaluate the efficacy of invariants through the proportion of equivalent mutants detected (i.e., mutations that yield syntactically different but semantically identical results), which is different from our goal of using them for distinguishing between effects from multithreading vs. error propagation.
Sahoo et al.~\cite{Sahoo2008} use likely invariants to detect hardware faults through software-level symptoms.
Their experiments show that their approach is able to identify over 95\% of hardware faults.
However, they focus only on range-based invariants (i.e., checking if values lie in a closed interval), significantly limiting the scope of the approach.
Further, they focus on hardware faults (i.e., single bit flips).
Lu et al.~\cite{Lu2006} develop a custom invariant extractor and utilize invariants to expose atomicity violations between thread interleavings.
In contrast to these papers, our paper explores the use of a broad set of likely invariants to trace the propagation of software run-time faults in multithreaded programs.


%% file: methodology.tex
\section{Methodology}
\label{sec:methodology}

We first formalize the EPA problem in Section~\ref{sec:system-model}.
Then, we provide an overview of our proposed solution in Section~\ref{sec:solution-overview}, followed by the development of \techname, the framework that implements our solution, in \Cref{sec:soba}.
We then present an example to show the applicability of \techname, followed by the evaluation metrics.

\input{system-model}

\subsection{Solution Overview}
\label{sec:solution-overview}

In our approach, we start with a set of golden runs $\Sigma$ and generate a set of likely invariants $F$ from them, before we inject a fault and run the program again.
The potentially faulty execution is then validated against the likely invariants.
Suppose $\sigma \in \Sigma$ denotes a golden run and $\sigma_f$ denotes a faulty execution of the program.
A likely invariant $f \in F$ is defined as a predicate over the set of reachable states $s_\sigma$ such that $f(s)$ is true for all $s \in s_\sigma$.
An execution $\sigma_f$ is said to deviate from the correct runs iff there is an invariant $f \in F$ such that $f(s)$ is false for some $s \in s_{\sigma_f}$.

\subsection{\techname: EPA Using Likely Invariants} 
\label{sec:soba}

\begin{figure*}
  \centering
  \includegraphics[keepaspectratio=true,width=0.75\linewidth]{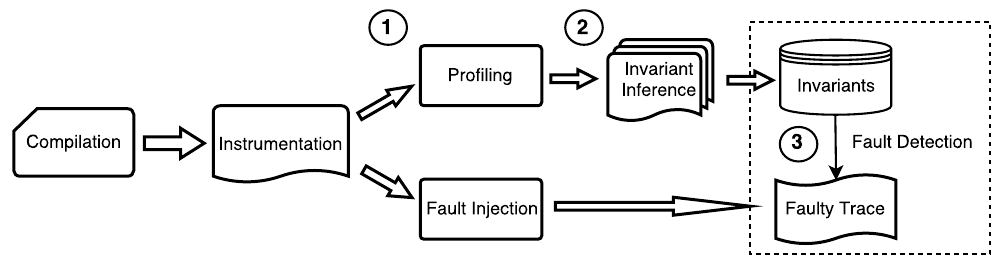}
  \caption{\techname: Invariant-based EPA Model}
  \label{fig:soba}
\end{figure*}

We now introduce \techname, a new EPA framework for multithreaded programs using dynamically inferred likely invariants.
\techname consists of three main modules, (1) program profiling, (2) invariant inference, and (3) fault detection.
Figure~\ref{fig:soba} overviews the EPA process using the \techname framework.

The \emph{profiling module} (label \ding{192}) is invoked at program compilation time, and instruments the tracing functions at the entry and exit points of every function in the program.
Tracing program values at function entry and exit points allows us to capture preconditions and postconditions of procedures, which broadly encapsulate its functionality.
A unique invocation nonce is also assigned to each pair of function entry and exit values, on a per thread basis.
The invocation nonce enables inferred invariants to associate exit values with entry values.
All of the traced values are accumulated in a trace file, which is then passed to the invariant inference module.

The \emph{invariant inference module} (label \ding{193}) examines the values in the trace file and generates likely invariants with a 100\% certainty, meaning that the invariants will never be falsified within the given trace file.
As discussed in Section~\ref{sec:solution-overview}, this stability across different runs is desired to keep the false positive rate low.
Therefore, programs must be checked to ensure that the set of likely invariants are stable for a given set of inputs.
Typically, for terminating programs, this problem can be remedied by using multiple profiling runs to generate the trace file.
Traces from multiple program runs can produce fewer invariants than single runs due to the heightened probability for falsification, but can also generate more invariants as larger traces offer higher statistical significance for previously neglected invariants.
Once the invariant inference module produces a stable set of invariants, the invariants can be deployed for validation against faulty traces (i.e., traces generated from faulty program runs).

Finally, the \emph{fault detection module} (label \ding{194}) parses and groups the invariants by their invoked functions.
These invariant groupings are stored in a hash map structure.
The faulty trace, which mirrors the format of the golden trace, is scanned line by line.
The fault detection module retrieves the corresponding invariant(s) from the hash map and validates the invariant(s) based on the faulty trace values.
The invariant violations are reported in a new file, which records the line number in the faulty trace, the function name, a flag indicating function entry or exit, and the violated invariant. 

\subsection{Example} 
\label{sec:example}

We outline an example of using \techname on the function shown in Figure~\ref{fig:func_example}.
As shown in Section~\ref{sec:motivating-example}, there is a potential for variance between golden traces of the same function, when executed with multiple threads.
Therefore, we demonstrate the application of \techname for EPA to circumvent the effects of non-determinism.

First, the \emph{profiling module} instruments the entry and exit points of this function while executing the function multiple times with a fixed input.
In this example, we pass a single fixed input, 4, to the function. 
$x$, is the only function argument variable traced at the function entry.
At the function exit, both the values of $x$ and the return values are traced.
Next, the \emph{invariant inference module} generates two sets of invariants at the entry and exit points respectively, using the trace files: $\{x = 4\}$, $\{x = 1\}$.
The entry invariant ($f$) specifies that all observed values of $x$ are equal to 4.
The exit invariant ($f'$) specifies that the final value of $x$ must be equal to 1.

Suppose a patch of the program incorrectly alters the boolean expression in line 3 to $a < 0$ (an easy mistake even by experienced programmers~\cite{Youngs1974, Ko2003}).
In fault injection, this bug can be simulated through a data corruption.
By inspecting the code, we observe that the bug leads to an erroneous change of $x$ at line 6.
The \emph{fault detection module} can detect this bug by validating the data trace of $x$ against the set of invariants, reporting the violation of ${f}'$.
We applied this mutation and found that \techname reports the violation of ${f}'$ in all of the faulty runs involving the same inputs on varying numbers of threads.

Violated invariants not only reveal the presence of faults, but also localize the source of faults. 
Since $f$ is retained, and ${f}'$ is violated, the fault must have occurred between the entry point and the exit point.
Note that these statements are not necessarily from the same function as other threads might have been interleaved with the function and might have modified the value of some of the variables.
Thus, an invariant based approach can avert the pernicious effects of thread variance.

\subsection{Evaluation Metrics}
\label{sec:evaluation-metrics}

We define the metrics which we use in later sections to evaluate the effectiveness of \techname.

{\bf Stability}: The invariant must hold across multiple fault-free executions of the programs targeted for injection with different numbers of threads for a given set of inputs.
This ensures a low false positive rate.

{\bf Coverage}: The invariants must provide high coverage for different types of faults, thereby ensuring a low false negative rate.
We define coverage of an invariant under a certain fault type as the rate of violations for the considered invariant under all injections for the considered fault type.

Suppose $T$ is the set of all faulty program traces, and $p$ is the number of violated invariants in a single trace.
Let $T_{p \geq 1}$ be a subset of $T$, denoting the set of program traces that violate at least one invariant. Then,
\begin{align*}
\mathrm{Fault\ Coverage} = \frac{|T_{p \geq 1}|}{|T|}
\end{align*}

{\bf Runtime Overhead}: 
The runtime overhead is a direct comparison between the time taken by \techname to complete the same task by golden run EPA.
We divide the overhead comparisons into two categories: setup overhead ratio ($S$), and fault detection overhead ratio ($D$).
The values of $S$ and $D$ are computed using \Cref{eq:setupOverhead,eq:detectionOverhead}, where the variable subscripts refer to the steps in EPA / IPA.

\begin{multicols}{2}
\noindent
\begin{equation}
  \label{eq:setupOverhead}
S = \frac{E_1}{I_1 + I_2}
\end{equation}
\begin{equation}
  \label{eq:detectionOverhead}
D = \frac{E_1 + E_3}{I_{1}/5 + I_3}
\end{equation}
\end{multicols}

In \techname, the one-time setup overhead for the fault injection experiments consists of golden run profiling ($I_1$) over 5 runs, and invariant generation ($I_2$). In EPA, only golden run profiling is performed ($E_1$), and there is no invariant generation step. 
Unlike the setup overhead which is a one-time cost, the fault detection overhead is incurred after every fault injection.
In \techname, this process consists of generating a single trace file ($I_{1}/5$) and executing the fault detection module ($I_3$).
In EPA, this process involves a line by line trace validation between golden and faulty runs ($E_3$).
We define the overhead ratios as time taken by EPA for setup/detection divided by that of \techname.


%% file: system-model.tex
\subsection{System Model}
\label{sec:system-model}

To systematically study the utility of likely invariants for EPA, we first introduce abstract models for sequential programs and parallel programs.
Using these models, we demonstrate that the most widely used approach to EPA for sequential programs is not applicable for multithreaded programs.
For brevity, we limit our models to terminating programs.
We do not consider this a restriction to our argument's generality, as most EPA techniques (like other experimental software assessment) evaluate correctness properties on a finite execution sequence of program statements.
For instance, a software test result is commonly evaluated against a \enquote{test oracle}, \emph{after the test execution has terminated}.
  If programs are strictly deterministic, i.e., their execution traces are always identical across repeated executions if no error occurs, then golden-run based EPA is sound irrespective of termination.
  However, if the execution traces of programs can deviate even for correct executions, e.g., because of multithreading, then golden-run based EPA becomes unsound for two reasons.
  First, the execution traces can capture different parts of the execution.
  Because termination is unknown, the length of an execution trace is an arbitrary choice and instructions that are executed outside of that arbitrarily chosen execution window could interleave with instructions from within that window in repeated executions.
  Second, the instructions within an arbitrarily chosen execution window could be permuted non-deterministically and mislead a direct comparison of executed instructions.
  Solving the first problem would require knowing all possible execution sequences, which contradicts non-termination in the general case.
  Therefore, our work focuses on the second problem.

  This restriction excludes certain types of programs from the application range for IPA.
  For example, server processes are usually designed to provide services continuously without termination, or even interruption.
  For these cases, neither EPA nor IPA can provide sound results and any improvement of IPA over EPA would be by coincidence.

\spara{EPA in Sequential Programs}
We define sequential programs by their control flow graphs (CFGs).
A CFG of a program $P$ is a directed graph $(V,E)$.
The set of vertices $V$ represents the program statements and the set of directed edges $E \subseteq \{(v_i,v_j) \in V \times V\}$ is defined such that $(v, v') \in E$ iff $v'$ is a \emph{possible direct successor} statement to $v$.
The relation $E$ follows directly from the sequence of statements in the program text and the programming language's semantics.
For every statement $v \in V$,  we define sets of predecessors $Pred(v) = \{v' \in V : (v',v) \in E\}$ and successors $Succ(v) = \{v' \in V : (v,v') \in E\}$.
A program has exactly one entry point $e$, and a single exit point $x$ such that $Pred(e) = \emptyset$ and $Succ(x) = \emptyset$. 
A program with multiple exit points can easily be modified to produce an equivalent program with one exit point.

We model an execution of a sequential program as a path in the CFG of the program.
Considering an execution $\sigma = e, v_1, \ldots, x$, we define a set of reachable states $s_\sigma$ such that each $s_i \in s_\sigma$ maps values to the program variables at each statement $v_i$ executed in $\sigma$.
For the sake simplicity, we write $s_\sigma(v_i)$ to refer to $s_i$.
The output of a program is solely determined by the provided input for a sequential program. 
The \emph{functional specification}  of a program relates all possible inputs to corresponding outputs.
A program execution whose output satisfies the functional specification is said to be \emph{correct}.
Any deviation of an execution from a correct execution with the same input is called an \emph{error}.
In a program execution, we refer to the sequence between an error and the last output-defining statement as \emph{error propagation}.

In EPA, faults are injected in the considered program to analyze their possible effects.
A \emph{fault injection} procedure adds, modifies, or removes a statement or its data, the \emph{injection point}, in the CFG.
Given a concrete input, the program is executed to obtain a correct execution, referred to as the \emph{golden run}. 
Next, a fault is injected and the program is executed again with the same input.
The obtained execution may deviate from the golden run as the code has been modified.
If so, the fault has been activated and resulted in an error, whose effects can be analyzed using the faulty execution trace.
Error propagation can be identified based on whether (1) the faulty execution $\sigma_f$ follows a different path in the CFG compared to the golden run $\sigma_g$ starting from the injection point $v_r$, OR (2) there exists a statement $v$ in $\sigma_f$ occurring after $v_r$, such that  $s_{\sigma_f}(v) \neq s_{\sigma_g}(v)$. 

\spara{EPA in Multithreaded Programs.} Multithreaded programs consist of multiple threads executing concurrently.
Each thread $T_i$ is modeled as a separate CFG $(V_i,E_i)$.
For a statement $v$ in $V_i$, we write $th(v)$ to refer to the thread $T_i$ executing it.
An execution of a concurrent program is a sequence of statements $\sigma = e_i, v_1, v_2, \ldots, x_j$ such that for any two statements $v$ and $v'$ with $v$ occurring before $v'$ in $\sigma$ and $th(v) = th(v')$,  $v \in Pred(v')$.
In other words, an execution is a linearization of partial orders of statements induced by the respective CFGs.

Due to the non-determinism of scheduling, different \enquote{equivalent} linearizations are possible.
Given the same input, two executions $\sigma = e_i,\ldots,x_i$ and $\sigma' = e_j,\ldots,x_j$ are said to be equivalent iff they deliver the same output, that is, $s_\sigma(x_i) = s_{\sigma'}(x_j)$.
Given a golden run $\sigma_r = \ldots, v_i, v_j, \ldots$, one can obtain an equivalent linearization $\sigma'_r = \ldots, v_j, v_i, \ldots$ by swapping adjacent statements $v_i$ and $v_j$, as long as the CFG order and synchronization mechanisms allow it and $s_{\sigma_r}(v_j) = s_{\sigma'_r}(v_i)$.
Successive swapping of such statements generates more possible linearizations that can characterize a correct execution \cite{mazurkiewicz1987}.
This is the reason why straightforward pairwise comparison of statements in the executions introduces unsoundness to golden run based EPA for multithreaded systems.
Thus, golden run based EPA may erroneously flag a deviation of observed equivalent linearizations due to to scheduler non-determinism.


%% file: evaluation.tex
\section{Experimental Evaluation} 
\label{sec:evaluation}

The goal of our experiments is to evaluate the effectiveness of the likely invariants derived by \techname in performing EPA.
As mentioned in \Cref{sec:solution-overview}, to be effective, a likely invariant should have two properties: (1) stability, and (2) coverage. 

To evaluate the stability, we execute the program multiple times, and measure the number of executions after which the invariant set stabilizes (Section~\ref{sec:rq_stability}).
We then measure the coverage provided by the invariants for different fault types by injecting faults into the program and checking whether any of the invariants are violated due to a fault (Section~\ref{sec:exp.coverage}).
We also group the invariants into different classes based on their structure, and measure the coverage provided by each class of invariants (Section~\ref{sec:exp.classes}).
Finally, we measure the performance overhead of the IPA and EPA approaches (Section~\ref{sec:exp.performance}).

\subsection{Research Questions}
\label{sec:exp.rq}

We ask the following research questions (RQ's) in our experimental evaluation.
\begin{itemize}
\item {\bf RQ1}: Is golden run based EPA a sound method to identify error propagation in multithreaded programs?
\item {\bf RQ2}: Do the invariants stabilize across multiple executions of the program?
\item {\bf RQ3}: What coverage of injected faults do the invariants provide for multithreaded programs and how does that coverage compare to singlethreaded EPA?
\item {\bf RQ4}: What is the coverage provided by invariants of a specific type/class, for different kinds of errors in the program?
\item {\bf RQ5}: What is the performance overhead of \techname compared to EPA?
\item {\bf RQ6}: Can stable invariants be generated at a lower confidence level?
\item {\bf RQ7}: Can stable invariants be generated at a finer program granularity?
\item {\bf RQ8}: Are program characteristics correlated with fault detection coverage of IPA?
\end{itemize}

\subsection{Experimental Setup}
\label{sec:exp.setup}

\techname\footnote{We have made \techname publicly available at \emph{\url{http://github.com/DependableSystemsLab/LLFI-IPA}}.} consists of three modules as shown in Figure~\ref{fig:soba}, namely the program profiling module, the invariant inference module, and the fault detection module. 
The program profiling module is implemented as a LLVM~\cite{llvm} compiler transformation pass, which is based on the instrumentation pass in the Udon tool \cite{Kusano2015}.
The invariant inference module utilizes \daikon \cite{Ernst2000}, since it is presently the most widely used tool for likely invariant generation.
Therefore, the primary function of the program profiling module is to produce a trace file in a \daikon-compatible format.
This involves some customized configurations in the LLVM compiler pass.
For simplicity of implementation, \techname only traces the values of function arguments belonging to primitive data types -- this is similar to what Udon~\cite{Kusano2015} does.
Lastly, the fault detection module consists of a single Python script and compares the values in the trace file with the derived invariants.

We evaluate the \techname framework using six representative multithreaded benchmarks that perform a wide variety of tasks: Quicksort, Blackscholes, Swaptions, Streamcluster, Nullhttpd, and Nbds.
These benchmarks range from roughly \num{300} to \num{3000} lines of code.
All benchmarks are implemented in C/C++, and use the POSIX threading library (i.e., \emph{pthreads}). We run all benchmarks using default program inputs that come with the benchmark suites.
Quicksort, as its name suggests, sorts a sequence of integers both sequentially and concurrently using the Quicksort algorithm, and returns a response code to denote success or failure. 
Blackscholes, Swaptions, Streamcluster are part of the PARSEC benchmark~\cite{PARSEC}, a suite of domain-diverse multithreaded programs of different workloads. 
Blackscholes is an application that solves the Black-Scholes partial differential equation, which prices a portfolio of European-style stock options.
Swaptions uses the Monte Carlo pricing algorithm to compute the prices of swaptions, a form of financial derivative.
Streamcluster is a web server application performing the online clustering problem with streaming data.
Nullhttpd is a small and efficient multithreaded web server for Linux and Windows~\cite{nullhttpd}.
Nbds~\cite{nbds} is an implementation of non-blocking data structures supporting concurrent key-value store transactions.
We choose these benchmarks to represent a wide variety of domains where multithreading is commonly applied.

We use LLFI~\cite{LLFI}, a LLVM based tool, to perform fault injections.
While LLFI was originally developed for hardware faults, it currently supports both software and hardware faults\footnote{Available at: https://github.com/DependableSystemsLab/LLFI}.
LLFI injects software faults into the program IR by modifying instructions or register values of the program at runtime.
We assume that faults are uniformly distributed throughout the program code.
Table~\ref{tab:faulttypes} describes how LLFI injects each software fault.
We consider only activated faults, i.e., those in which modified instructions are executed or modified data is read by the program, when reporting coverage.

\subsection{Fault Model}
\label{sec:fault-model}

In this paper, we consider the following 6 software faults listed in Table~\ref{tab:faulttypes}: data corruptions, file I/O buffer overflows, buffer overflows (involving) malloc, function call corruptions, invalid pointers and race conditions.
The chosen fault types are conceptually closer to errors that occur at run-time than faults in programs (i.e., software bugs), which are typically used in compile-time injection approaches such as G-SWFIT \cite{Duraes2006} or SAFE \cite{Natella2013}.
For instance, buffer overflows can result from either miscalculated offsets into a data structure or from insufficient memory allocations.
The reason for this choice lies in our goal to study error propagation.
By focusing on \emph{effects} (e.g., the overflow), we do not need to simulate all their individual root causes separately, which would require more experiment time 
and result in lower propagation probabilities, because these root causes do not necessarily lead to wrong program executions.

Data corruption is a generic fault type that can capture a wide variety of errors due to logical errors (e.g., the example in Section~\ref{sec:methodology}), and implementation bugs (e.g., integer overflows, uninitialized variables).
The buffer overflow fault categories can occur due to common bugs in C/C++ programs where array and pointer bounds are not checked.
We distinguish between file I/O-related buffer overflows and other buffer overflows as the former can lead to security vulnerabilities.
Function call corruptions can occur when one passes the wrong parameters to a function, and represents incorrect invocation of functions i.e., interface errors.
Invalid pointers can arise due to errors in pointer arithmetic, or due to the use of pointers after freeing them, i.e., use-after-free bugs.
Finally, race conditions occur due to locks not being acquired or acquired incorrectly, and with at least one of the threads performing a write to shared data. 
We limit ourselves to six fault modes to keep the number of experiments tractable -- all six faults are supported by LLFI~\cite{RaiyatAliabadi2016}.
Note that the fault types above are broader than those covered by traditional mutation testing, in that they involve corruption of the program state beyond simple syntactic changes. 

The six chosen software faults represent common bugs~\cite{V2005} that are difficult to capture through unit or regression tests, and have been used in prior work to emulate software faults~\cite{Hsueh1997, Ghosh1998, Jeffrey2008}. 
Memory-related bugs such as data corruptions, buffer overflows, function call corruptions, and invalid pointers are particularly hard to debug as the observed program failures may not be clearly associated with these faults~\cite{Jeffrey2008}.
This is attributed to two main reasons.
First, a single memory bug can propagate to other locations, causing multiple memory bugs. 
Eventually, program failures may be attributed to memory bugs that are often distant from the root memory bug. 
Secondly, different types of memory bugs may lead to one another through propagation. 
For example, a data corruption can lead to a buffer overflow, which may cause a function call corruption.
Such scenarios entail significant effort to determine their root causes.
Error testing in unit and regression tests aims to validate the program's exception handling rather than exhaustively testing for the presence of memory bugs.
Like memory-related bugs, race conditions often lead to effects that cannot be detected by simple failure detectors.
Therefore, EPA is particularly useful for analyzing the effects of these bugs, which makes them an important target in our evaluation.

\begin{table}
    \caption{Description of faults injected using LLFI}
    \label{tab:faulttypes}
\begin{center}
    \begin{tabular}{|l|l|}
    \hline
    \textbf{Fault Type} & \textbf{LLFI Implementation} \\ \hline
    Data Corruption & Randomly flips a single bit in an arbitrary data value in the program  \\ \hline
    File I/O Buffer Overflow &  Randomly increases the \emph{size} in \emph{fread} and \emph{fwrite} operations \\ \hline
    Buffer Overflow Malloc & Under allocates malloc and calloc to emulate overflowing the allocated buffers \\ \hline
    Function Call Corruption & Randomly corrupts the source register (i.e., parameter) of a function call \\ \hline
    Invalid Pointer & Randomly corrupts the returned pointer from malloc and calloc\\ \hline
    Race Condition & Replaces a lock of a mutex in the program with a fake mutex\\ \hline
    \end{tabular}
    \end{center}
\end{table}

\subsection{RQ1: Golden Run Variance}
\label{sec:rq0}

\begin{figure}[tbh]
  \centering
  \includegraphics[keepaspectratio=true,width=0.75\columnwidth]{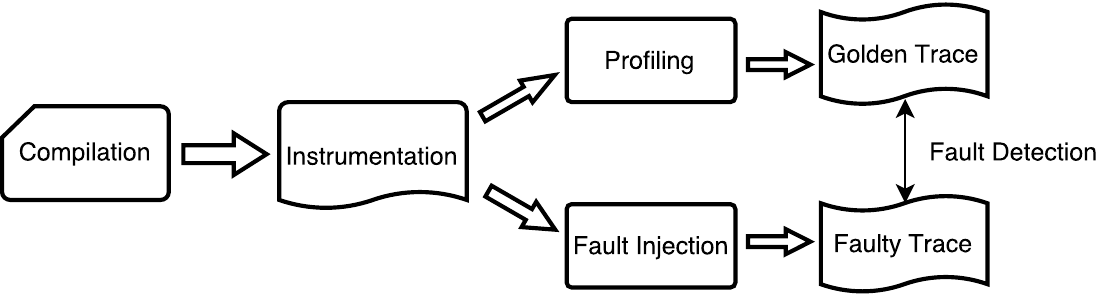}
  \caption{Golden run based EPA}
  \label{fig:existing_model}
\end{figure}

We conduct golden trace analysis (the traditional EPA model) over the benchmark applications (see \Cref{sec:exp.setup}), by varying the number of threads for each program. 
To conduct EPA following the traditional EPA model shown in Figure~\ref{fig:existing_model}, the application is compiled and instrumented to invoke a tracing function at every LLVM IR instruction.
Hence, each line in a trace file represents an instruction identifier and its corresponding data value in the program.
A golden trace of the original program instructions is generated in a process known as \emph{profiling}.
Then, a fault is injected into the program and a trace of the modified program instructions is produced.
Finally, EPA is performed by comparing the golden and faulty traces line by line.
Discrepancies between the two traces will reveal how faults propagate through the program execution paths.

\begin{figure}[tbh]
  \centering
  \includegraphics[keepaspectratio=true,width=0.6\columnwidth]{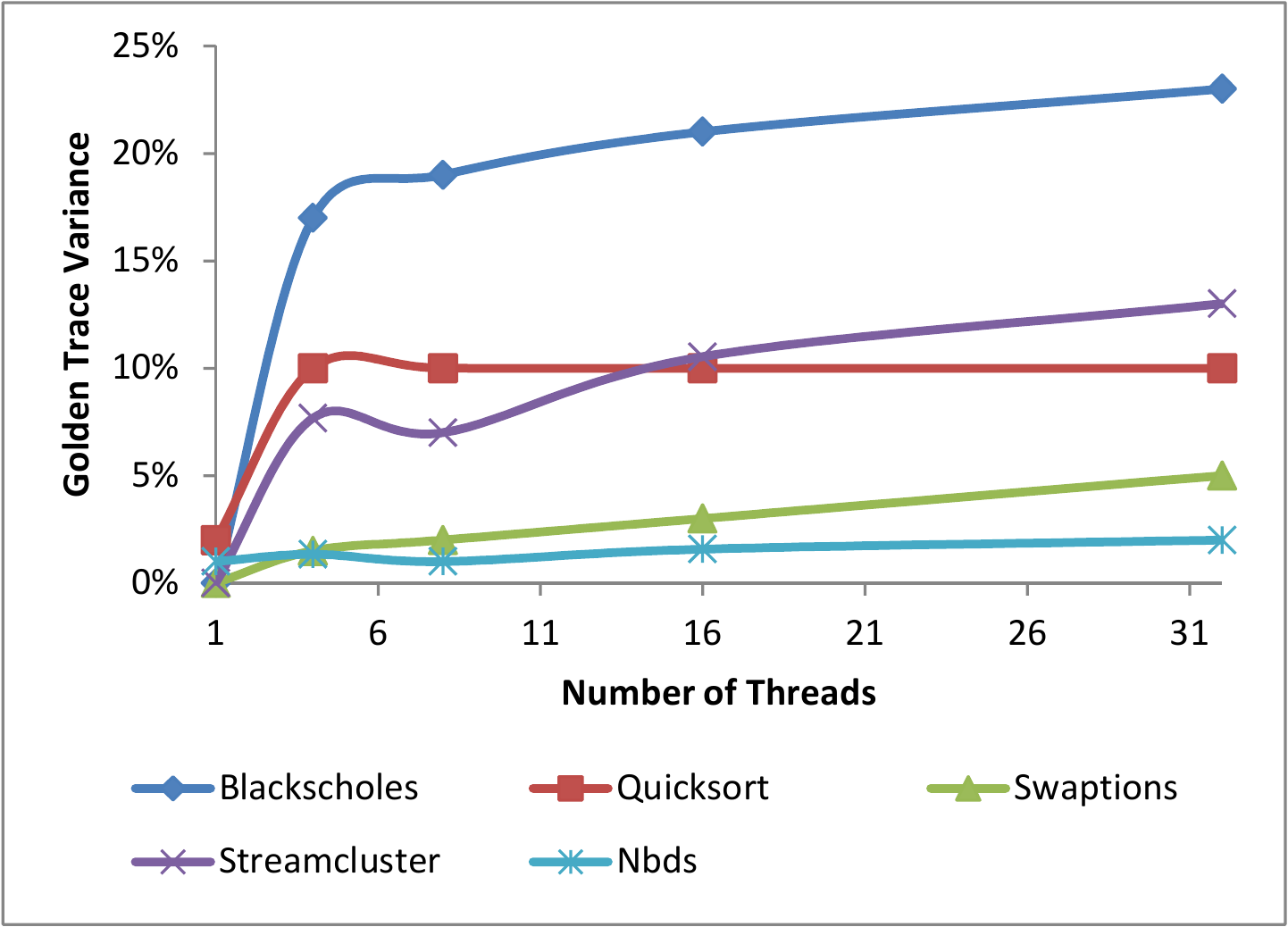}
  \caption{Average variance between golden run traces}
  \label{fig:variance}
\end{figure}

We collect golden runs over all benchmark programs except \emph{Nullhttpd}~\footnote{This experiment was not conducted on Nullhttpd since the thread number was not externally configurable beyond single- vs. multithreading.}, running them with a single thread, \num{4} threads, \num{8} threads, \num{16} threads, and \num{32} threads respectively.
We find considerable variance between the golden traces upon running the applications with different numbers of threads using the same input, which obviously does not indicate error propagation.
Variance is measured by taking the proportion of line conflicts between two trace files relative to the total number of lines in a single trace file (i.e., proportion of dynamic instructions with different values).

Figure~\ref{fig:variance} depicts the average variances between \num{5} golden traces of each application, executed with different numbers of threads.
The variance between the golden runs is \SI{10}{\percent} on average due to multithreading non-determinism.

Note that it is possible to use traditional EPA for the deterministic portions of the program (cf. \cite{Lemos}).
However, it is non-trivial to identify the deterministic portions \emph{a~priori}, as these depend both on the number of threads and the inputs given to the program.
Therefore, traditional methods for EPA cannot be used in a multithreaded context. 

\begin{obs}
  \label{obs:variance}
  If a multithreaded program is repeatedly executed with the same input, the golden runs extracted from these executions differ from each other, which renders traditional golden run based EPA unsound.
\end{obs}

\subsection{RQ2: Stability}
\label{sec:rq_stability}

\begin{figure}[htb]
  \centering
  \includegraphics[keepaspectratio=true,width=0.75\columnwidth]{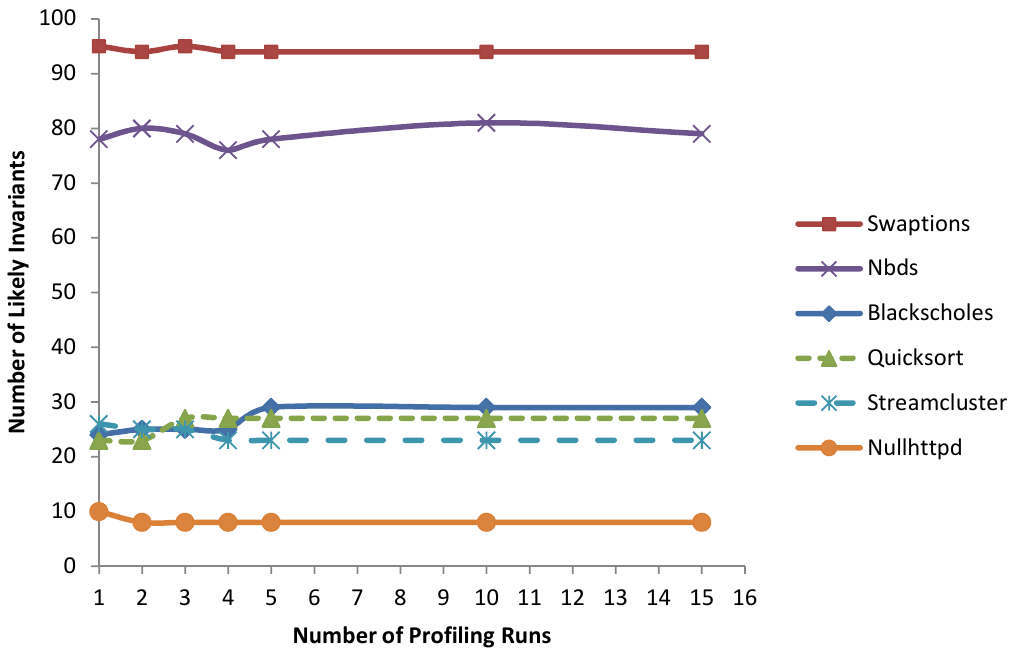}
  \caption{Number of invariants generated from varying numbers of profiling runs for six benchmark applications}
  \label{fig:stability}
\end{figure}

For likely invariants to improve EPA's resilience to effects from multithreading non-determinism, the generated invariants must be \emph{reproducible} among repeated program executions.
In this experiment, we evaluate the stability of the set of dynamically generated invariants across execution reiterations.
Let $n$ denote the number of execution recurrences.
Each application begins with $n=1$ to produce a trace file, which is then delivered to the invariant inference module.
The invariant inference module returns a single set of invariants.
This process is repeated with $n=2,3,4,5,10,15$, resulting in a family of sets of invariants.
The number of invariants obtained at each $n$ value is reported in Figure~\ref{fig:stability}.
In all of our sample applications, we observe a convergence of likely invariants by $n=5$.
We also verified manually that the invariant sets match when the invariants converge, i.e., the invariants derived are the same after \num{5} executions.

Table~\ref{tab:invariantcounts} shows the counts of inferred invariants in our sample applications.
These are shown only for the stable invariants.
We find that there is roughly one invariant for every \numrange[range-phrase = --]{10}{100} lines of source code, with the sole exception of Nullhttpd.
Few invariants were inferred from Nullhttpd as many of its functions were parameterless. 
This ratio is captured by the invariant density, $\rho$, which represents the number of invariants per lines of code.
The invariant counts show that stable invariants can be inferred from multithreaded programs, when repeatedly executed with the same inputs. 

\begin{table*}
    \caption{Invariant counts and classification (refer to Table~\ref{tab:invariantclass}) of \techname's generated invariants}
    \label{tab:invariantcounts}
	\begin{center}
    \begin{tabular}{|l|c|c|c|c|c c c c c c c c c|}
    \hline
    \multirow{2}{*}{\textbf{Benchmark}} & \multirow{2}{*}{\textbf{LOC}} & \multirow{2}{*}{\textbf{Functions}} & \multirow{2}{*}{\textbf{Invariants}} & \multirow{2}{*}{\textbf{$\rho$ (\%)}} & \multicolumn{9}{c|}{\textbf{Invariant Classes}}\\
    \cline{6-14} & & & & &
    \textbf{A} & \textbf{B} & \textbf{C} & \textbf{D} & \textbf{E} & \textbf{F} & \textbf{G} & \textbf{H} & \textbf{Other}\\ \hline
    Quicksort & 330 & 9 & 27 & 8.2 & 3 & - & - & - & 1 & 1 & 16 & 6 & -\\
    Blackscholes & 526 & 5 & 29 & 5.5 & - & - & - & - & 3 & - & 15 & 11 & -\\
    Streamcluster & 1580 & 11 & 23 & 1.5 & 1 & - & - & - & - & - & 14 & 6 & 2\\
    Swaptions & 1635 & 14 & 94 & 5.7 & 7 & 4 & 3 & 1 & 4 & 4 & 59 & 11 & 1\\
    Nullhttpd & 2500  & 20 & 8 & 0.3 & - & - & - & 2 & - & - & 4 & 2 & -\\
    Nbds & 3158  & 27 & 80 & 2.5 & - & - & - & - & 4 & - & 36 & 39 & 1\\ \hline
    \end{tabular}
    \end{center}
\end{table*}

\begin{obs}
  \label{obs:stability}
  If a multithreaded program is repeatedly executed with the same input, the likely invariants generated from these executions stabilize within five executions.
\end{obs}

For our coverage assessment in the following section, we consider only the stable invariants, or those invariants that hold across all observed executions (in our experiments). This allows us to minimize the number of false positives and obtain conservative lower bounds on the coverage. 

\subsection{RQ3: Coverage}
\label{sec:exp.coverage}

As we showed in the previous sections, using invariants instead of golden run based comparisons, 
we were able to improve the soundness of EPA for multithreaded applications, i.e., minimize false positives.
An important question is, whether we also miss true positives in the process of reducing false positives, i.e., if the likelihood of false negatives is increased for invariant based EPA.
To answer this question, we perform \num{1000} fault injections of each fault type in Table~\ref{tab:faulttypes}, one per run, on the benchmark applications. 
Subsequently, we compare the faulty program traces against the set of inferred invariants.
If any of the likely invariants was violated due to the injected fault, we label the run as a successful detection. 

\begin{figure*}
  \subfloat[\label{fig:quicksortFI}]{\includegraphics[width=0.45\linewidth]{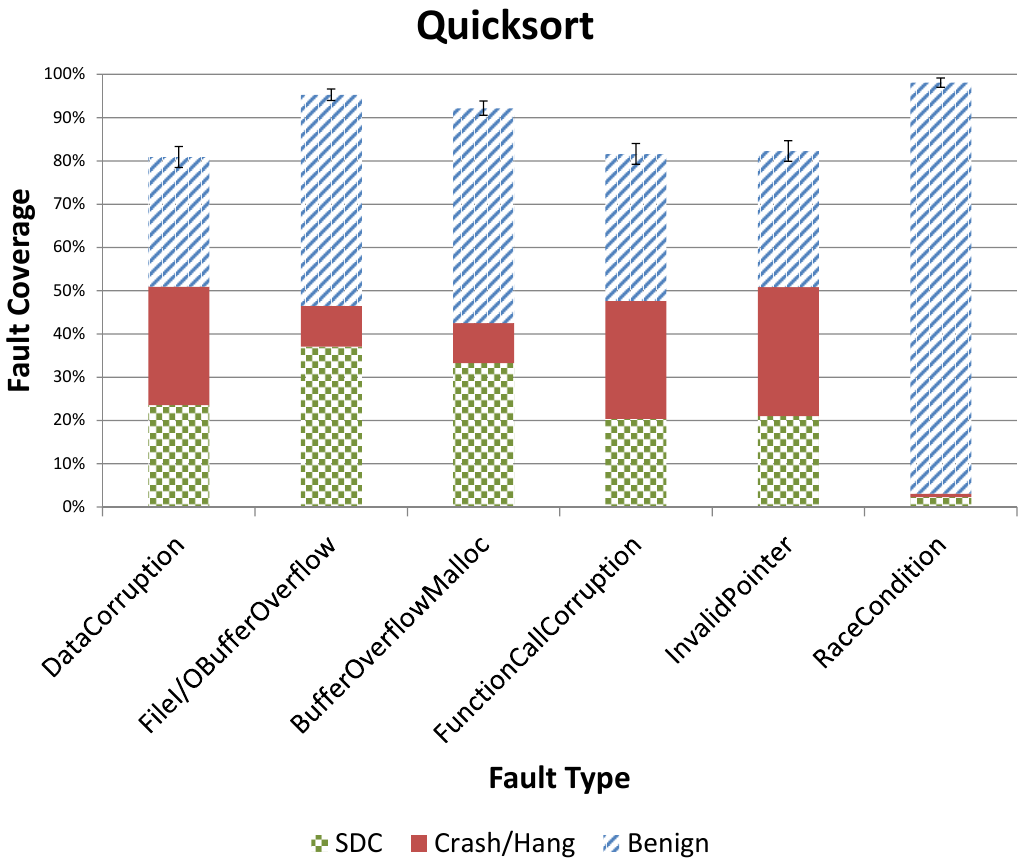}\hspace{0.5em}} \hfill
  \subfloat[\label{fig:blackscholesFI}]{\includegraphics[width=0.45\linewidth]{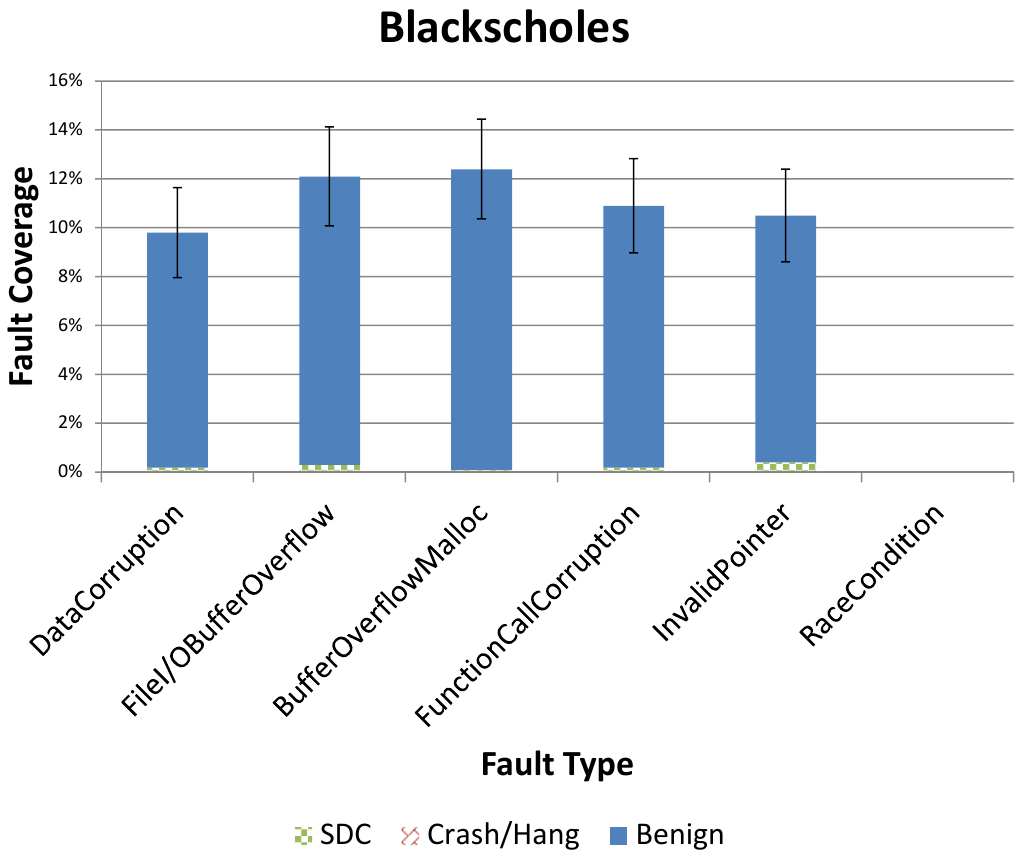}\hspace{0.1em}} \\

  \subfloat[\label{fig:swaptionsFI}]{\includegraphics[width=0.45\linewidth]{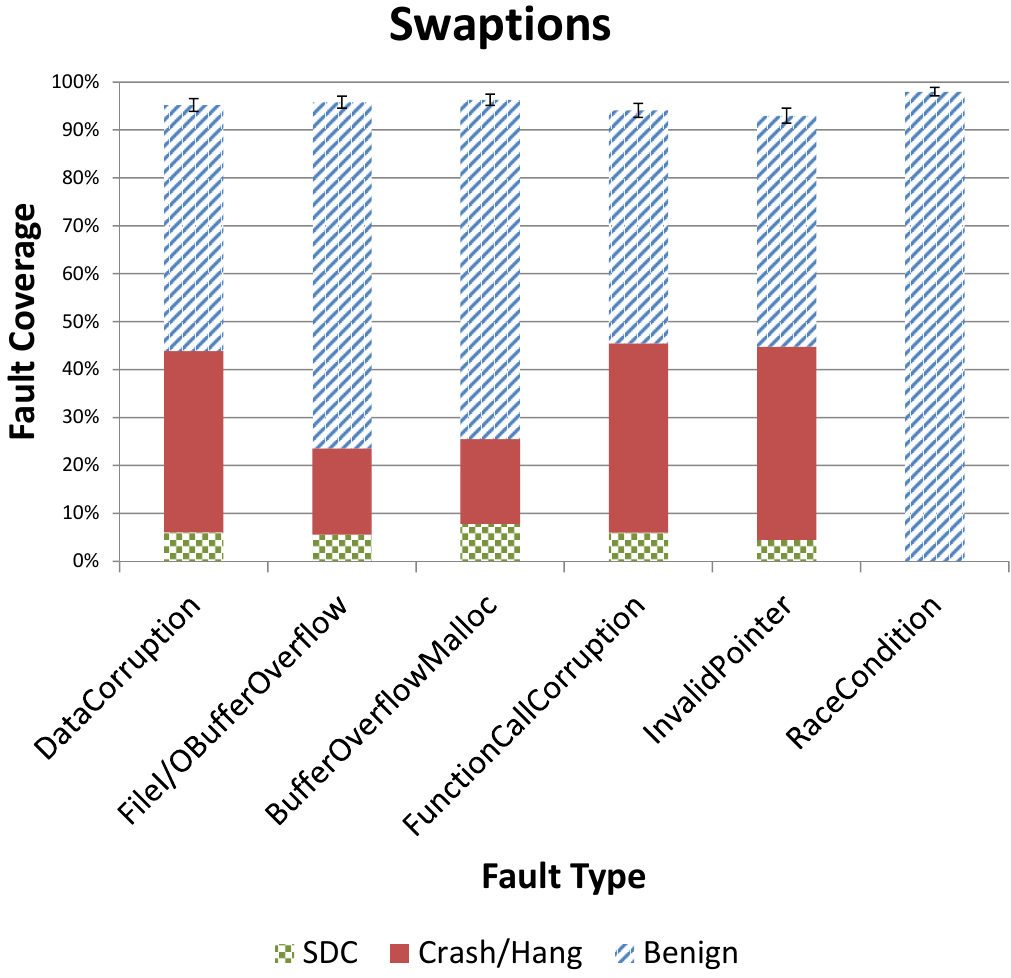}\hspace{0.5em}} \hfill
  \subfloat[\label{fig:streamclusterFI}]{\includegraphics[width=0.45\linewidth]{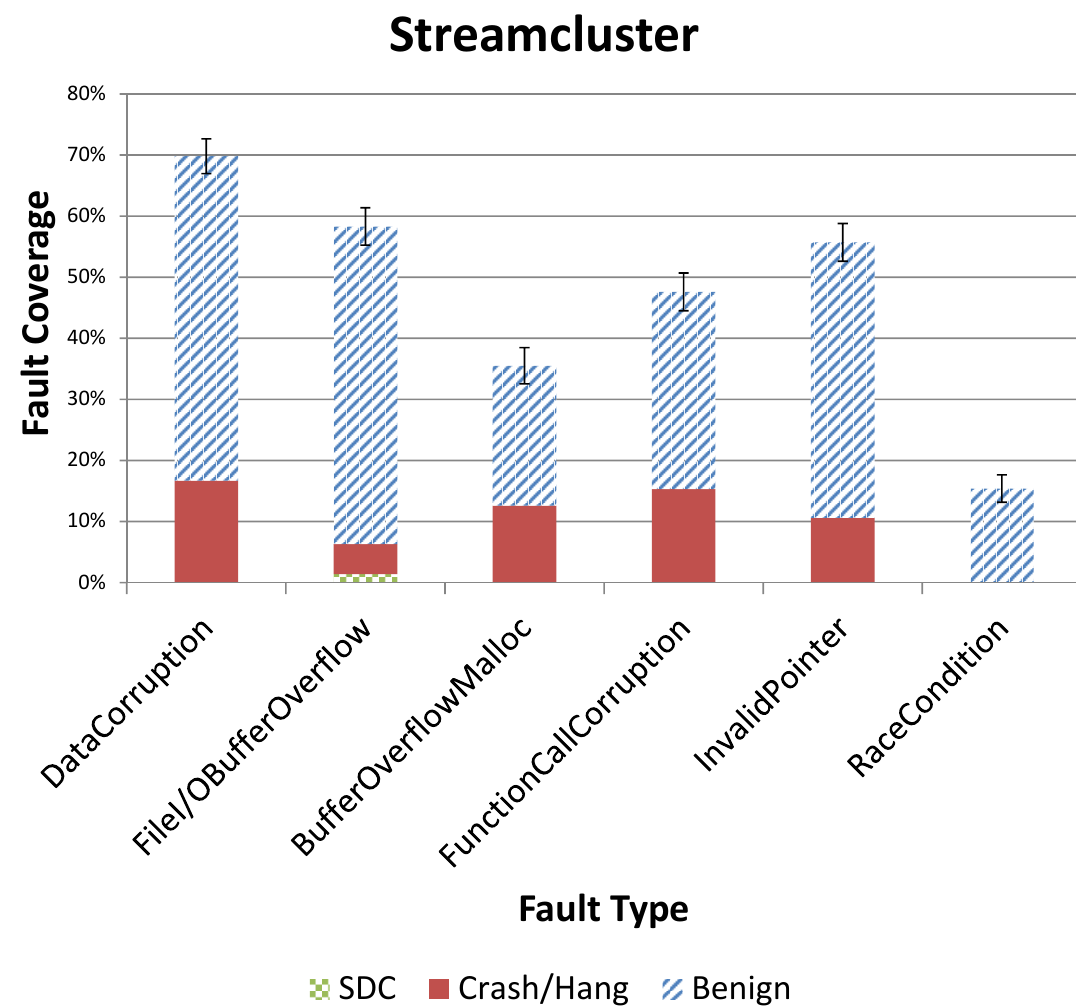}\hspace{0.1em}} \\

  \subfloat[\label{fig:nullhttpdFI}]{\includegraphics[width=0.45\linewidth]{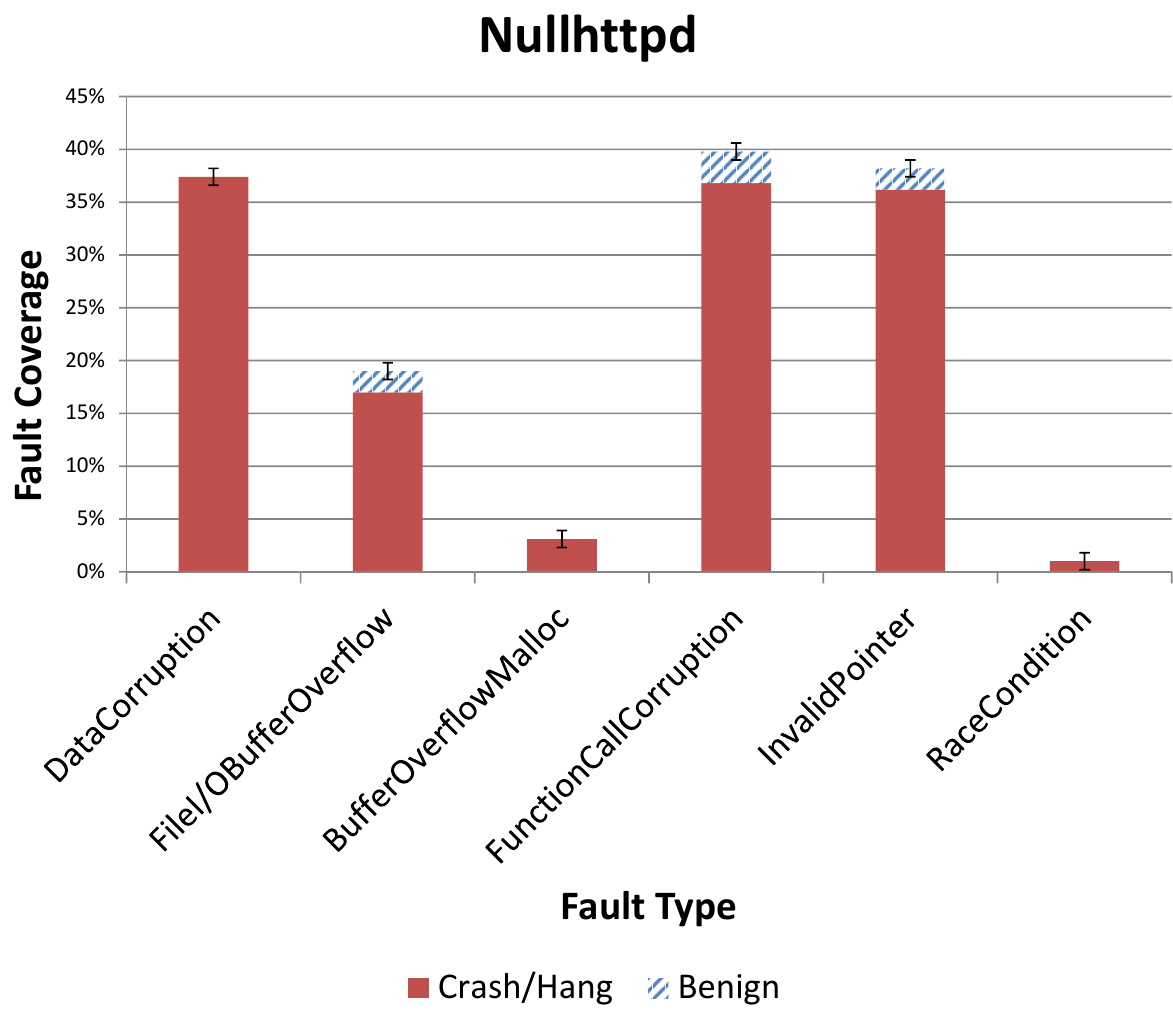}\hspace{0.5em}} \hfill
  \subfloat[\label{fig:nbdsFI}]{\includegraphics[width=0.45\linewidth]{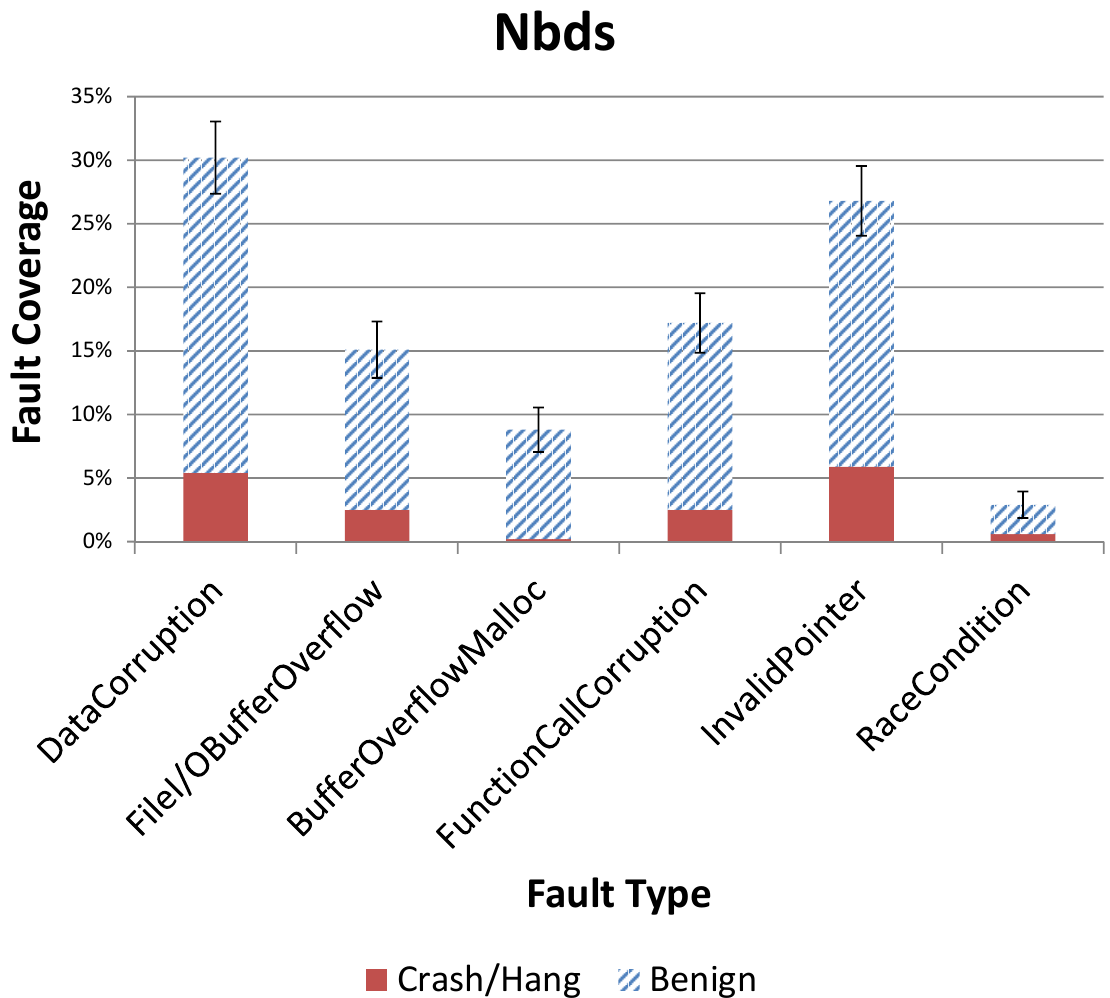}\hspace{0.1em}} \\

\caption{Proportion of \num{1000} faulty runs that violate at least one invariant}
\end{figure*}

The fault coverages for each application are shown in \Cref{fig:quicksortFI,fig:swaptionsFI,fig:blackscholesFI,fig:streamclusterFI,fig:nbdsFI,fig:nullhttpdFI}.
The error bounds denote the \SI{95}{\percent} confidence intervals of the reported fault coverages. 
The figures show the fault coverage for different fault types divided into three failure modes, as seen in similar fault injection experiments~\cite{LLFI}: Benign, Crash/Hang and Silent Data Corruption (SDC).
Benign indicates faulty program runs with no observable deviations in the final program output.
Faults may still propagate through internal functions without manifesting into an observable output deviation.
Crash/Hang signifies faulty runs that either terminate with exceptions or time out.
SDC specifies faulty runs that terminate normally but produce program outputs that deviate from the golden run (i.e., incorrect outputs).
SDCs are often the most important failure modes, as they are much harder to detect than crashes.

We find that the fault coverage provided by the invariants varies widely across applications, from \SIrange[range-phrase = --]{90}{97}{\percent} for \emph{Swaptions}, to \SIrange[range-phrase = --]{10}{15}{\percent} for \emph{Blackscholes}.
This variation occurs due to fluctuations in two factors: Invariant densities ($\rho$), and invariant relevance (i.e., ability of the invariant to detect faults).
Quicksort and Swaptions have higher invariant densities at \SI{8.2}{\percent} and \SI{5.7}{\percent} respectively.
However, invariant density does not express the relevance of the invariants to fault detection.
The sets of invariants for Quicksort and Swaptions both contain a number of invariants involving computation data, while Blackscholes is dominated by invariants on local environment variables.
Computation data is more likely to be passed inter-procedurally, which increases the likelihood of fault detection.
In contrast, local environment variables rarely carry beyond the scope of functions.
Consider the case where a variable is corrupted at the function exit.
If no invariants exist on that variable at the function exit, the fault would not be captured.
However, the prospect of fault detection increases if the value is passed to subsequent functions, which may have invariants checking it.

Further, there is considerable variation across different fault types and their consequences on the benchmark applications.
For example, in \emph{Streamcluster}, the coverage for race conditions is only about \SI{15}{\percent}, while it is \SI{70}{\percent} for data corruption errors.
In other benchmarks (e.g., Quicksort), the situation is reversed, with race conditions having the highest coverage (\SI{97}{\percent}), while data corruption errors have the lowest coverage (\SI{80}{\percent}).
Data corruption errors directly affect the data as data operand bits are randomly flipped.
On the contrary, the effects of race conditions can be difficult to predict as they are dependent on the implementation of locking patterns in the threading library.
In this case, race conditions cause Quicksort and Swaptions to violate (some) invariants, yet minimal effects are observed in other benchmarks.

Across all applications, the benign errors constitute a majority of fault outcomes (\SI{73}{\percent} on average), followed by Crash/Hang (\SI{22}{\percent}) and SDCs (\SI{5}{\percent}).
We do not measure SDCs in Nullhttpd and Nbds since the applications return either a successful response code or a failure message.
We find that benign errors exhibit the highest fault coverage overall.
Although benign errors are typically neglected in EPA, benign fault coverage shows that invariants can track benign faults before they are masked.
This may be important to find latent bugs in the program.
On the contrary, Crash/Hang are the most blatant failures. 
Nullhttpd has the highest rate of Crash/Hang fault coverage among the benchmarks. 
We find that a set of initialization invariants are violated whenever the web server fails to load.
Finally, SDCs are typically the least commonly observed failure outcomes across applications, and consequently have the least coverage. 
Quicksort has the highest rates of SDC error detection among all the applications. This is because it contains many inequalities, and a single negated inequality can impact the final ordering of values. Correspondingly, many of the invariants in Quicksort consist of inequality conditions and ordering constraints that are sensitive to such value deviations, and hence yield high coverage.
In RQ8, we study the correlation of the coverage of IPA with different program metrics.

While the high false positive rates render golden run based EPA infeasible for multithreaded programs, it does constitute a gold standard for singlethreaded programs.
Therefore, we investigate how the coverage of invariant based EPA for multithreaded programs compares to golden run based EPA for singlethreaded programs.
For this purpose we reconfigure and recompile our target programs to execute with a single thread and repeat the injection campaigns for those.
For Nbds we were not able to achieve a strictly singlethreaded configuration without massively altering the code, such that a valid comparison would be questionable.
Therefore, we exclude Nbds from our comparison between singlethreaded and multithreaded EPA.

\begin{figure}[!htbp]
  \centering
  \includegraphics[keepaspectratio=true,width=\textwidth]{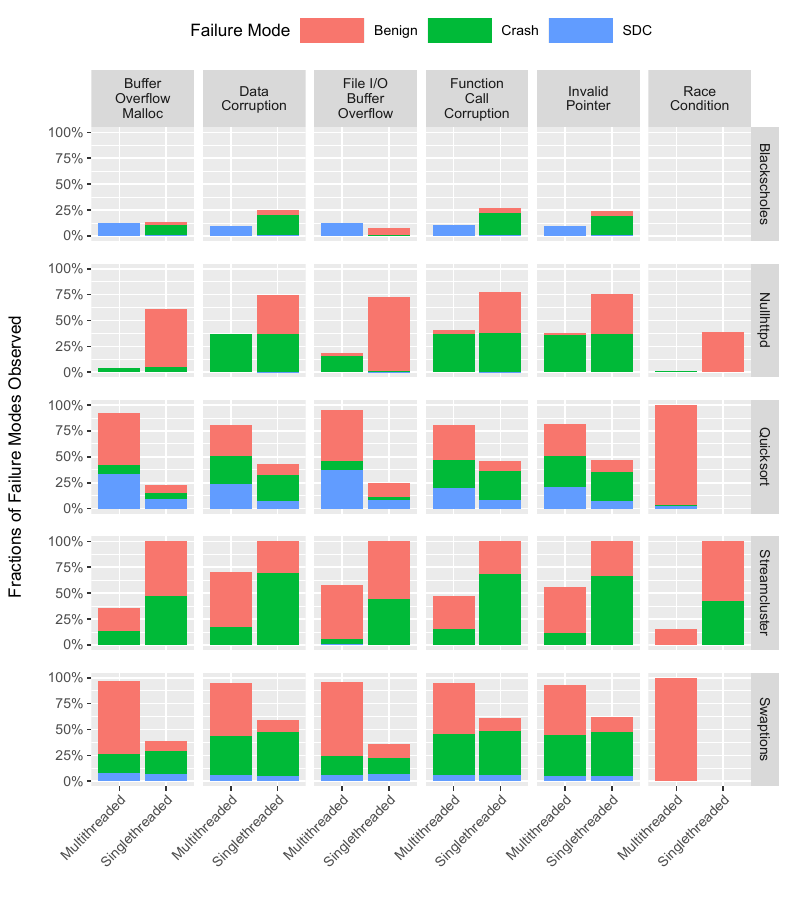}
  \caption{EPA result differences between invariant based EPA for multithreaded programs and golden run based EPA for singlethreaded programs}
  \label{fig:multisingle}
\end{figure}

\Cref{fig:multisingle} shows the obtained coverage per target program and fault model in comparison to the coverage for multithreaded EPA.
We first observe that the overall coverage is higher for singlethreaded EPA in the case of Blackscholes, Nullhttpd, and Streamcluster, while it is lower for Quicksort and Swaptions.
A naive interpretation of this observation would be that IPA's false negatives dominate for Blackscholes, Nullhttpd, and Streamcluster, whereas its false positives dominate for Quicksort and Swaptions.
However, a more thorough investigation of the failure mode distribution refutes this conclusion.
For Blackscholes, Quicksort, and Streamcluster the failure mode distributions obviously differ significantly between singlethreaded and multithreaded fault injection experiments.
SDC failures are much more prevalent in the multithreaded experiments compared to singlethreaded experiments with Blackscholes and Quicksort, whereas there is a strong difference in the number of crashes for Streamcluster.
However, for these failure modes (SDC and Crashes), the differences in coverage cannot be attributed to false positives or negatives by IPA's propagation analysis, because these failure modes prove that propagation has taken place: its effect has caused the program to output wrong data or to terminate abnormally.

A manual inspection of the experiment logs confirmed the observed differences for SDCs and crashes.
This means that errors propagate very differently depending on whether programs are executed with many or with a single thread.
In consequence, a direct coverage comparison between singlethreaded and multithreaded EPA techniques cannot yield meaningful results.
To the best of our knowledge this is the first empirical evidence that fault injection results are affected by whether a program uses one or many threads.
This implies that the resilience and error propagation characteristics of multithreaded programs cannot be inferred from fault injection experiments on singlethreaded configurations of those programs.

This highlights an interesting dichotomy.
  On the one hand, the multithreaded and singlethreaded variants of the programs implement the same function, i.e., their externally observable behavior is identical modulo execution time.
  On the other hand, the structure and the execution characteristics of these variants differ significantly.
  Our results show that these significant differences, although they do not affect functionality, do have an influence on the programs' failure characteristics.

These findings stress the need for dedicated propagation analyses for multithreaded programs such as IPA.

\begin{obs}
  \label{obs:faultcoverage}
  Errors propagate differently in multithreaded and singlethreaded configurations of the same program.
  Hence, dedicated propagation analysis techniques are required for both cases.
  If faults are injected in a multithreaded application, their effects are indicated by violations of likely invariants generated from fault-free multithreaded executions of that application. The coverage provided depends both on the application and the type of faults injected. 
\end{obs}

\subsection{RQ4: Invariant Classification}
\label{sec:exp.classes}

During the automated inference of likely invariants, we observed that many invariants have a similar structure. 
For example, some invariants involve inequalities, while others involve set membership and ordering. 
This observation leads us to ask whether differences in structure of the invariants correlate with differences in the respective invariants' effectiveness for EPA.
The result can help discover what constitutes a good invariant for EPA.

To study this effect, we first classify the invariants into eight different classes based on their structure and then consider the coverage of the invariant classes.
The classes are: Array-equality, elementwise-initialization, elementwise, initialization, inequality conditions, multi-value, order, return-value invariants.
Table~\ref{tab:invariantclass} provides a brief description of each invariant class~\footnote{The rightmost column of Table~\ref{tab:invariantcounts} shows the number of invariants per class in each benchmark.}.
The invariants are classified exclusively, without overlap between classes.
A small number of invariants did not fall into any of these eight classes -- we ignore them for this study. 

We calculate the coverage of an invariant class as the fraction of fault injection runs that violate at least \emph{one of the invariants} in that class.
For example, if an invariant class $I$ has two invariants $I_1$ and $I_2$, and $S_1$ and $S_2$ are the sets of fault injection runs that result in violation of the invariants $I_1$ and $I_2$ respectively, then the coverage of the invariant class $I$ is given by $(| S_1 \cup S_2 | )/N$, where $N$ is the total number of fault injection runs that had activated faults.

Table~\ref{tab:invariantcounts} shows the number of invariants that occur in different classes for the five applications. 
Due to space constraints, we only show the results for Swaptions, which provides the highest diversity of invariant classes.
\Cref{tab:swaptionsByClass} shows the results of the fault injection experiment for Swaptions, grouped by invariant classes.

\begin{table}
    \caption{Description of Invariant Classes}
    \label{tab:invariantclass}
	\begin{center}
	\scalebox{1}{
    \begin{tabular}{|c|p{1.6cm}|p{5.3cm}|}
    \hline
    \multicolumn{2}{|c|}{\textbf{Invariant Class}} & \textbf{Description}\\ \hline
    \textbf{A} & Array-equality & Equality condition on every element of an array \\ \hline
    \textbf{B} & Elementwise-initialization & Initial values of array elements \\ \hline
    \textbf{C} & Elementwise & Condition on the elements of an array \\ \hline
    \textbf{D} & Initialization & Invariants that associate post-conditions to pre-conditions \\ \hline
    \textbf{E} & Multi-value & Variable value must match exactly one element of a set \\ \hline
    \textbf{F} & Order & Array is sorted in ascending or descending\\ \hline
    \textbf{G} & Relational conditions & Invariants involving both equalities and inequalities \\ \hline
    \textbf{H} & Return-value & Invariants involving the return value of a function\\ \hline
    \end{tabular}
    }
    \end{center}
\end{table}

\begin{table}
\centering
  \caption{Classification of violated invariants from 1000 faulty Swaptions runs and their coverage}
    \label{tab:swaptionsByClass}
    \begin{tabular}{|l|l| c c c c c|}
    \hline
    \multirow{2}{*}{\textbf{Fault Type}} & \multirow{2}{*}{\textbf{Failure}} & \multicolumn{5}{c|}{\textbf{Invariant Classes (\%)}}\\
    \cline{3-7} & &
    \textbf{A} & \textbf{B} & \textbf{C} & \textbf{D} & \textbf{H}\\ \hline
    
    \multirow{3}{*}{DataCorruption} & SDC & 3 & 3 & 6 & 1 & 1\\
    & Crash & - & - & 38 & - & -\\
    & Benign & 26 & 26 & 51 & - & -\\ \hline
    \multirow{3}{*}{FileI/OBufferOverflow} & SDC & 3 & 3 & 6 & 1 & 1\\
    & Crash & 1 & 1 & 18 & - & -\\
    & Benign & 42 & 41 & 72 & - & -\\ \hline
    \multirow{3}{*}{BufferOverflowMalloc} & SDC & 4 & 4 & 8 & 1 & 1\\
    & Crash & - & - & 18 & - & -\\
    & Benign & 42 & 42 & 71 & - & -\\ \hline
    \multirow{3}{*}{FunctionCallCorruption} & SDC & 2 & 2 & 6 & 1 & 1\\
    & Crash & - & - & 40 & - & -\\
    & Benign & 26 & 26 & 49 & - & -\\ \hline
    \multirow{3}{*}{InvalidPointer} & SDC & 2 & 2 & 5 & - & -\\
    & Crash & - & - & 40 & - & -\\
    & Benign & 28 & 28 & 48 & - & -\\ \hline
    \multirow{3}{*}{RaceCondition} & SDC & - & - & - & - & -\\
    & Crash & - & - & - & - & -\\
    & Benign & 58 & 58 & 70 & - & -\\ \hline
    
    \end{tabular}
\end{table}

The elementwise invariants have the highest fault coverage overall in Swaptions (Table~\ref{tab:swaptionsByClass}).
Elementwise invariants correspond to predicates on individual elements of an array.
Swaptions stores its dataset in an array, which is passed back and forth between its functions.
As a result, a number of array element constraints arise.
Elementwise invariants offer a marginally higher fault detection rate for SDCs compared to the other invariant classes.

The fact that elementwise invariants have proven particularly effective is rooted in the central role of one particular data structure in Swaption's implementation.
We demonstrate an example from Swaptions in Figure~\ref{fig:swaptions_order_example}.
\texttt{funcA()} takes multiple inputs, including two 2D arrays of equal dimensions, \texttt{ppdFactors} and \texttt{ppdFacBreak}.
\texttt{funcA()} calculates \texttt{ppdFactors} using \texttt{ppdFacBreak} and another variable via \texttt{funcB()}.
Due to the calculation in \texttt{funcB()} and the value ranges for its parameters, which derive from stock option properties, it is close to impossible that this calculation yields a value equal or smaller than \texttt{ppdFacBreak}.
Daikon generates a post-conditional invariant that captures this relation.
It infers an elementwise (Class C) invariant, where every element of \texttt{ppdFactors} is always greater than the element located at the same indices in \texttt{ppdFacBreak}.
After performing our fault injection experiments, we found that this was one of the invariants that yielded high fault coverage.
If even a single element of the array(s) violates the invariant after a fault injection run, a potential fault is reported.

Conversely, we have found this invariant class to be of far less importance for other target programs, for which we found other classes to be more effective based on the programs' application logic and implementation.
For Quicksort, for instance, the order in which numbers are returned in each recursive step are of central importance.
Unsurprisingly, we found order and return-value invariants to collectively yield high SDC fault detection for that program.

\begin{figure}[htb]\
\begin{Verbatim}[frame=single,fontsize={\normalsize},numbers=left,numbersep=5pt,xleftmargin=10pt]
int funcA(ppdFactors[][], ppdFacBreak[][], ...) {
	for(i=0; i<=M; ++i){
		for(j=0; j<=N; ++j) {
			ppdFactors[i][j] = funcB(...);
			...
//Class C invariant: ppdFactors[][] > ppdFacBreak[][]
}
\end{Verbatim}
\caption{Example function in Swaptions with a Class C "Elementwise" post-condition invariant.}
\label{fig:swaptions_order_example}
\end{figure}

\begin{obs}
  The  coverage of invariants for an application differs across different classes of likely invariants generated from fault-free multithreaded executions of the application.
\end{obs}

\subsection{RQ5: Performance Evaluation}
\label{sec:exp.performance}

We evaluate the performance of \techname by comparing each step shown in Figure~\ref{fig:soba}, described in Section~\ref{sec:soba}, to its equivalent in EPA.
Table~\ref{tab:perfComp} exhibits the average durations for each step, measured in seconds, each averaged over 5 runs.
No faults were injected in this experiment as we wanted to obtain the worst case performance overheads (i.e., when the application executes to completion). 

We find that \techname induces a \SIrange[range-phrase = --]{2}{90}{\percent} setup overhead over EPA, while \techname is 2.7$\times$ to as much as 151$\times$ faster than EPA for fault detection, depending on the benchmark.
This is because EPA traces the program execution after each point, while \techname only checks for consistency of invariants at function entries and exits.

Thus, \techname incurs a slightly higher setup overhead compared to EPA for the programs in our study, but has substantially lower fault detection overheads.
Examining the setup overheads in relation to the complexity of the programs (\Cref{tab:invariantcounts}), 
we notice that the programs with high setup overheads
\emph{tend to have} few lines of code, which hints at an effect from the cost of EPA.
  EPA requires every single instruction in the program to be logged during a profiling run.
  Therefore, programs with many lines of code have a higher setup overhead with EPA.  Therefore,
  the setup overhead for IPA becomes smaller compared to the overhead that EPA entails for these programs.
Since the fault detection overhead is incurred on each run (numbering thousands in a typical fault injection experiment), 
it is more important than the setup overhead, which is a one-time cost.

\begin{table}
    \caption{\techname vs EPA performance measured in seconds for step numbers 1, 2 and 3 (refer to Figure~\ref{fig:soba}).}
    \label{tab:perfComp}
	\begin{center}
    \begin{tabular}{|l|c c c|c c|l|l|}
    \hline
    \multirow{2}{*}{\textbf{Benchmark}} & \multicolumn{3}{c|}{\textbf{\techname}} & \multicolumn{2}{c|}{\textbf{EPA}} & \multirow{2}{*}{\textbf{S$\times$}} & \multirow{2}{*}{\textbf{D$\times$}}\\
    \cline{2-6} & \textbf{1} & \textbf{2} & \textbf{3} &
    \textbf{1} & \textbf{3} & & \\ \hline
    Quicksort & 5.5 & 5.4 & 0.4 & 1.1 & 1.4 & 0.1 & 2.7\\
    Blackscholes & 5.5 & 8 & 0.5 & 4.1 & 72 & 0.29 & 72\\
    Streamcluster & 5.5 & 7.3 & 0.7 & 4.1 & 52.2 & 0.31 & 44\\
    Swaptions & 9.5 & 8.9 & 1.1 & 18.1 & $>$300 & 0.96 & 151\\
    Nullhttpd & 5.6 & 4.9 & 0.3 & 3.6 & 22.1 & 0.32 & 27\\
    Nbds & 32 & 18.3 & 7.4 & 49.2 & 28.3 & 0.98 & 7.3\\ \hline
    \end{tabular}
    \end{center}
\end{table}

\begin{obs}
  \label{obs:performance}
  \techname incurs a higher setup overhead compared to EPA, but has significantly lower fault detection overhead.
\end{obs}

\subsection{RQ6: Inferring Invariants at a Lower Confidence}
\label{sec:confidence}

In Section~\ref{sec:exp.setup}, we only consider invariants inferred from the data traces at the 99\% Daikon confidence level. 
In this section, we conduct an analysis of whether it is possible to infer more stable invariants at lower confidence levels.
The confidence represents the probability that an invariant is not inferred randomly (i.e., with one sample out of thousands of traces)~\cite{Ernst2000}.
The availability of more stable invariants at lower confidence levels may offer higher fault coverage.

We refer to the Daikon confidence of invariants as confidence rather than their statistical confidence.
Each class of invariant has a different measure of confidence with respect to the samples provided~\cite{Ernst2000}.
For instance, a relational condition invariant such as $a < b$ has a confidence of $1 - {\frac{1}{2}}^{n}$ where $n$ is the number of samples in the data trace. 
However, an invariant that is falsified within the data trace file always has a confidence equal to 0.

We conduct an experiment where we infer invariants at various confidence levels by adjusting the \enquote{conf\_limit} option in Daikon.
Table~\ref{tab:confidence} shows the number of invariants inferred at the 99\%, 80\% and 60\% levels.
We take the number of invariants at the 99\% confidence as the baseline. 
Of the four benchmarks shown, there is a modest increase of invariants, ranging from a 0\% to 4\% increase at the 80\% confidence and 0\% to 10\% increase at the 60\% level.
The number of invariants for Swaptions and Nullhttpd did not change with the confidence levels. 

\begin{table}
    \caption{Number of invariants inferred per confidence level.}
    \label{tab:confidence}
	\begin{center}
	\scalebox{1}{
    \begin{tabular}{|c|c|c|c|c|}
    \hline
    \textbf{Conf.} & \textbf{Quicksort} & \textbf{Blackscholes} & \textbf{Streamcluster} & \textbf{Nbds} \\ \hline 
    99\% & 25 & 24 & 29 & 73\\
    80\% & 26 & 24 & 29 & 76\\
    60\% & 28 & 25 & 32 & 79\\ \hline
    \end{tabular}
    }
    \end{center}
\end{table}

\begin{obs}
  The number of invariants do not significantly increase when the invariants are inferred at a lower confidence. 
\end{obs}

\subsection{RQ7: Inferring Invariants at a Finer Granularity}
\label{sec:granularity}

In Sections~\ref{sec:exp.setup} through \ref{sec:confidence}, \techname infers invariants at the function level.
We explore the feasibility of inferring invariants at a finer program granularity, i.e., at the basic block level.
The purpose of this experiment is to determine whether we can increase fault coverage through finer program granularity between invariants.
For instance, consider the function in Figure~\ref{fig:blackscholes} from the Blackscholes benchmark.
This function can be divided into 6 basic blocks as shown in Figure~\ref{fig:example_cfg} -- we infer invariants at the beginning of each block.
The invariants at the beginning of the first basic block (lines 1-7) and the second basic block (lines 8-9) respectively are: \lstinline|{InputX > 0}| and \lstinline|{InputX < 0}|.
If the condition in line 7 is erroneously changed to \lstinline|InputX > 0.0| (same type of fault previously described in Section~\ref{sec:example}), the fault detection module can invalidate the faulty program trace through the \lstinline|{InputX < 0}| invariant.
Using basic block invariants, the fault is detected in the second basic block rather than the end of the function.
\techname may not always detect faults by relying solely on invariants at the function entry and exit points.
A fault may lead to an observable error in the body of a function, where a function variable is corrupted. 
In this example, \lstinline|InputX| may be overwritten in the \lstinline|computeNPrimeX()| function on line 13 and the function exit invariants may not detect the fault.

To enable \techname to infer invariants at basic blocks, we modify the instrumentation step of \techname in Figure~\ref{fig:soba} by adding a LLVM pass to instrument programs at the entry point of each basic block.
Like the function instrumentation pass in \techname, the modified pass traces the values of function argument variables rather than intermediate variables.
We also change the fault detection script to validate the program trace values against basic block invariants instead of function invariants.
All other modules in the \techname workflow remain the same in this experiment.

\begin{figure}[tbh]
  \centering
  \includegraphics[keepaspectratio=true,width=0.75\linewidth]{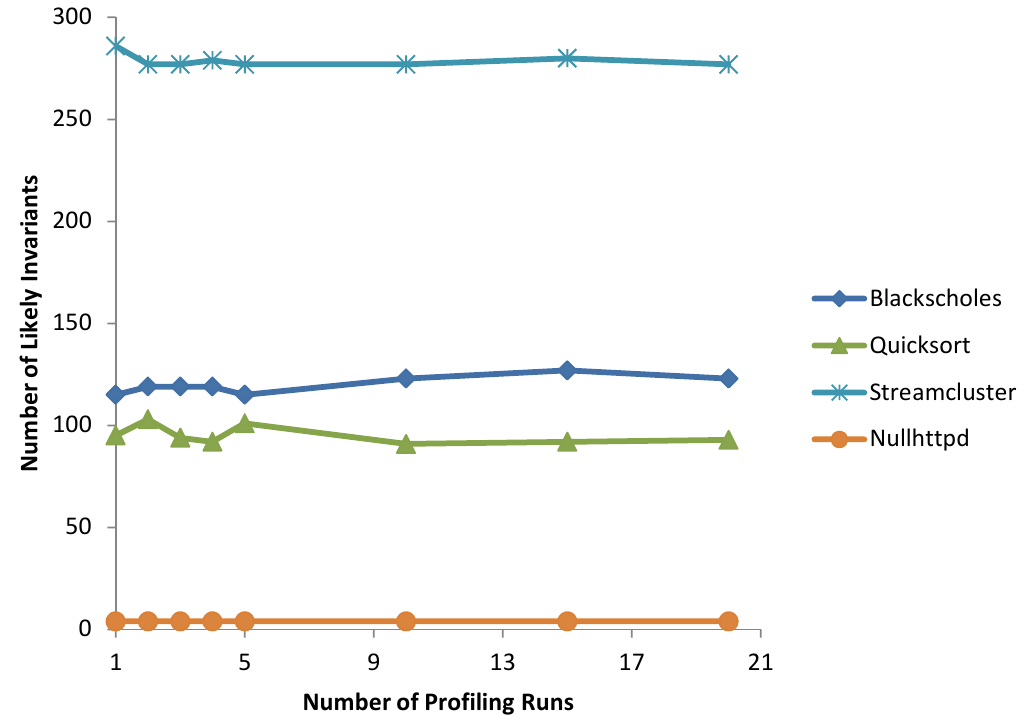}
  \caption{Number of invariants generated from varying numbers of profiling runs at the basic block level.}
  \label{fig:BB_all}
\end{figure}

\begin{figure}[tbh]
  \centering
  \includegraphics[keepaspectratio=true,width=0.75\linewidth]{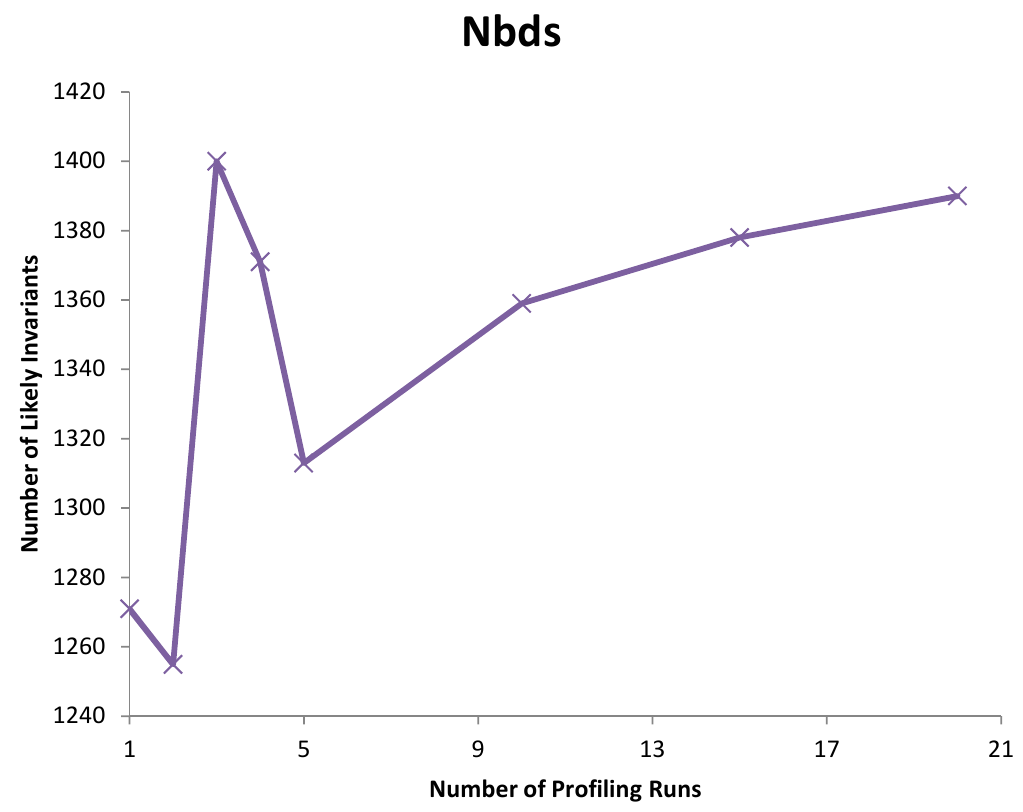}
  \caption{Number of invariants generated from varying numbers of profiling runs at the basic block level.}
  \label{fig:BB_nbds}
\end{figure}

We repeat the experiment in Section~\ref{sec:rq_stability} to determine whether the generated invariants stabilize within a fixed number of runs.
In Figure~\ref{fig:BB_all}, we observe that the invariants from four benchmark programs do not stabilize within 20 runs.
Additionally, Nbds exhibits large fluctuations in inferred invariants as seen in Figure~\ref{fig:BB_nbds}.
We display Nbds separately as the number of its invariants significantly exceeds the range of other benchmark programs.
However, Swaptions and Nullhttpd have no fluctuations in their inferred invariants.
We observe that the invariant fluctuations are greater in benchmarks whose functions were divided into a large number of basic blocks.
For example, functions in Swaptions were divided into 4 basic blocks on average while many functions in Nbds were divided into 8 or more basic blocks.  
The finer division of functions into basic blocks also signify the increased presence of branching.
The values of function variables likely fluctuate around the branch conditions.
Based on these observations, invariants inferred at the basic block granularity are not suitable for fault detection.
The numbers of invariants differ significantly between profiling runs and the set of invariants used as a test oracle does not stabilize within the range of 20 runs.

\begin{figure}[htb]\
\begin{Verbatim}[frame=single,fontsize={\normalsize},numbers=left,numbersep=5pt,xleftmargin=10pt]
fptype CNDF ( fptype InputX ) {
    int sign; // BB 1

    fptype OutputX;

    // Check for negative value of InputX
    if (InputX < 0.0) {
        InputX = -InputX; // BB 2
        sign = 1;
    } else // BB 3
        sign = 0;
    
    OutputX = computeNPrimeX(InputX); // BB 4
    
    if (sign) {
        OutputX = 1.0 - OutputX; // BB 5
    }
    
    return OutputX; // BB 6
}
\end{Verbatim}
\caption{Example function in the Blackscholes application}
\label{fig:blackscholes}
\end{figure}

\begin{figure}[tbh]
  \centering		
  \includegraphics[keepaspectratio=true, scale=1]{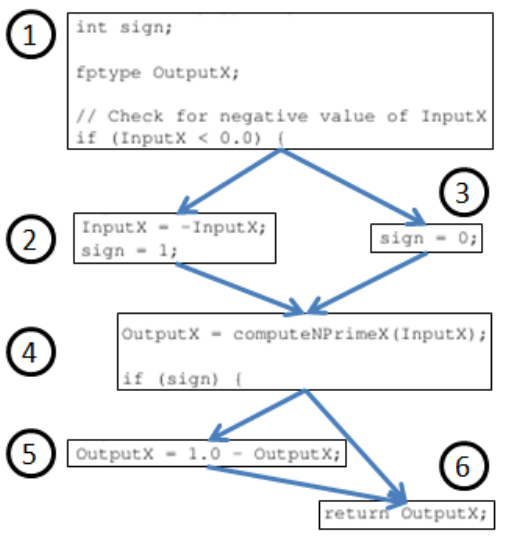}		
  \caption{Control flow graph with labelled basic blocks of the function shown in Figure~\ref{fig:blackscholes}}		
  \label{fig:example_cfg}		
\end{figure}

\begin{obs}
  Likely invariants inferred at the basic block granularity under repeated program executions with the same input do not stabilize within twenty executions. 
	Therefore, these invariants are not suitable for fault detection.
\end{obs}

\subsection{RQ8: Are program characteristics correlated with fault detection coverage of IPA?}

  To better understand why IPA performed so differently for the programs in our study, we conducted a multiple correlation test between the observed coverages for different fault types and various program metrics.
  We used \texttt{cqmetrics} \cite{Spinellis2016}, a tool that calculates program metrics for C programs by statically analyzing the program's source code.
  We collected a total of 111 metrics across our targeted programs using \texttt{cqmetrics}.
  However, many of these metrics  do not affect the LLVM IR of the programs that LLFI injections work on and, thus, cannot affect the results of our experiments,  
  e.g., metrics related to spaces/indentation or comments. So we excluded these metrics.
  After this initial filtering, we were left with 25 variables, four of which were constantly \num{0} for our programs (the number of \texttt{goto}, \texttt{union}, \texttt{noalias}, and \texttt{signed} keywords).
  The remaining 21 metrics used in our analysis are listed in \Cref{tab:metrics}.
  
  We conducted a pairwise Spearman correlation test for each of the 21 metrics with IPA's fault coverage for the different fault models in our study. 
  We also studied the correlations among the metrics themselves to control for confounding.
  \Cref{tab:corr} shows the metrics that have statistically significant correlations with IPA's fault coverage across the benchmarks.

  We find that almost none of the metrics exhibit a statistically significant correlation with IPA's coverage, with the exception of one metric, namely \emph{the number of structs}. 
  The number of structs in a program has a strong negative correlation across all fault models, except for race conditions, which has generally low coverage across all programs.
  This is because invariant inference engines, such as Daikon, do not support invariants on complex data types, such as structs or objects.
  Therefore, if errors propagate via these data types, they cannot be detected, because no invariants are generated.

  We also observed statistically significant negative correlations between other source code metrics and the coverage for buffer overflow faults, as shown in \Cref{tab:corr}.
  However, these metrics also exhibit a strong positive correlation with the number of structs for the programs in our study and, hence, we assume their correlation with coverage to be an effect of their correlation with struct counts.
 Therefore, we can conclude that the single metric that has a significant impact on IPA's coverage in our study is the number of structs in the program.

\begin{obs}
 For programs that make fewer use of complex data types, like structs or objects, IPA achieves a higher coverage. However, the coverage is not correlated with any other program metric.
\end{obs}

\begin{table*}[ht]
  \centering
  \caption{Code metrics used in correlation analysis with IPA fault coverage.\label{tab:metrics}}
  \rowcolors{1}{}{lightgray}
  \begin{tabular}{p{5.5cm}rrrrrr}
    \toprule
    Metric & blackscholes & nbds & nullhttpd & quicksort & streamcluster & swaptions \\
    \midrule
    Number of statements & 654.00 & 2068.00 & 998.00 & 109.00 & 869.00 & 462.00 \\
    Statement nesting mean & 2.00 & 16.89 & 5.37 & 0.34 & 1.67 & 5.23 \\
    Statement nesting median & 1.00 & 13.50 & 3.00 & 0.00 & 1.00 & 1.00 \\
    Statement nesting std.\,dev & 3.32 & 18.35 & 6.09 & 0.61 & 1.99 & 7.34 \\
    Number declarations with internal (\texttt{static}) visibility & 0.00 & 106.00 & 2.00 & 2.00 & 7.00 & 3.00 \\
    Number of \texttt{const} keywords & 7.00 & 114.00 & 17.00 & 8.00 & 16.00 & 14.00 \\
    Number of \texttt{enum} keywords & 0.00 & 3.00 & 0.00 & 0.00 & 0.00 & 0.00 \\
    Number of \texttt{inline} keywords & 0.00 & 14.00 & 0.00 & 0.00 & 1.00 & 0.00 \\
    Number of \texttt{register} keywords & 0.00 & 0.00 & 0.00 & 0.00 & 0.00 & 1.00 \\
    Number of \texttt{restrict} keywords & 0.00 & 5.00 & 0.00 & 0.00 & 0.00 & 0.00 \\
    Number of \texttt{struct} keywords & 7.00 & 74.00 & 50.00 & 6.00 & 6.00 & 3.00 \\
    Number of \texttt{unsigned} keywords & 6.00 & 36.00 & 30.00 & 1.00 & 6.00 & 0.00 \\
    Number of \texttt{void} keywords & 16.00 & 311.00 & 67.00 & 24.00 & 24.00 & 20.00 \\
    Number of \texttt{volatile} keywords & 0.00 & 22.00 & 0.00 & 0.00 & 5.00 & 0.00 \\
    Number of \texttt{typedef} keywords & 4.00 & 71.00 & 12.00 & 0.00 & 4.00 & 1.00 \\
    Halstead complexity mean & 3148.00 & 12683.39 & 16623.61 & 386.35 & 1173.81 & 10340.98 \\
    Halstead complexity median & 2921.65 & 10060.84 & 15724.94 & 114.71 & 705.33 & 8531.95 \\
    Halstead complexity std.\,dev. & 1391.37 & 9708.58 & 7424.94 & 545.76 & 1532.78 & 4460.43 \\
    Cyclomatic complexity mean & 15.90 & 130.23 & 87.18 & 2.91 & 11.36 & 50.85 \\
    Cyclomatic complexity median & 15.00 & 111.50 & 77.50 & 2.00 & 6.50 & 42.00 \\
    Cyclomatic complexity std.\,dev. & 2.36 & 95.84 & 34.53 & 2.31 & 12.95 & 21.69 \\
    \bottomrule
  \end{tabular}
\end{table*}

\begin{table}[ht]
  \centering
    \caption{ Spearman correlation coefficients between program metrics and IPA's fault coverage. Statistically significant correlations ($\alpha=0.05$) are highlighted with grey background color.
	\emph{Rows without any statistically significant correlations are omitted from the table.} }\label{tab:corr}
      \begin{adjustbox}{max width=\columnwidth}
  \begin{tabular}{lrrrrrr}
    \toprule
    \multirow{3}{*}{Metric} & Buffer   & \multirow{2}{*}{Data}       & File I/O  & Function   & \multirow{2}{*}{Invalid} & \multirow{2}{*}{Race}\\
                            & Overflow & \multirow{2}{*}{Corruption} & Buffer    & Call       & \multirow{2}{*}{Pointer} & \multirow{2}{*}{Condition}\\
                            & Malloc   &                             & Overflow  & Corruption &                          & \\
      \midrule
      nstatement & \cellcolor{lightgray}{-0.83} & -0.60 & -0.60 & -0.60 & -0.60 & -0.58 \\
      nstruct & \cellcolor{lightgray}{-0.93} & \cellcolor{lightgray}{-0.81} & \cellcolor{lightgray}{-0.81} & \cellcolor{lightgray}{-0.81} & \cellcolor{lightgray}{-0.81} & -0.72 \\
      nunsigned & \cellcolor{lightgray}{-0.93} & -0.75 & -0.75 & -0.75 & -0.75 & -0.68 \\
      ntypedef & \cellcolor{lightgray}{-0.87} & -0.70 & -0.70 & -0.70 & -0.70 & -0.68 \\
      \bottomrule
  \end{tabular}
  \end{adjustbox}
\end{table}


%% file: discussion.tex
\section{Discussion}
\label{sec:discussion}

We first present the implications of our results, and then the threats to the validity of our study. 

\subsection{Implications}

In this paper, we address the question of whether likely invariants derived by automated techniques can be used for EPA in multithreaded programs. 
EPA requires stable invariants, which provide high coverage for different types of faults. 
We find that the invariants stabilize within a few executions of the program.
However, their coverage is highly dependent on the application.
For some applications, the achieved coverage is high (\SIrange{80}{90}{\percent}), while for other applications, it is quite low (\SI{10}{\percent} or less).
The type of inferred invariants is another factor for consideration.
In Table~\ref{tab:invariantcounts}, relational invariants (Type G) are predominant in all benchmarks.
Conversely, as seen in both \Cref{tab:swaptionsByClass}, their fault coverages are low.
This suggests that existing invariants derived by automated tools such as \daikon~\cite{Ernst2000} may not be sufficient to ensure high fault coverage across applications.

Furthermore, the coverage provided by the invariants depends on the specific fault that is injected, e.g., race conditions.
Finally, most of the invariants provide coverage for benign failures and crashes, both of which are much more numerous than SDCs.
However, SDCs are an important concern in practice, as they can result in catastrophic failures, and likely invariants do not currently provide high coverage for SDCs.
Improving the coverage of likely invariants for SDCs is a direction for future work. 

We further study the effect of invariant structure on fault coverage by grouping the invariants into different categories.
Similar to the prior experiments in Section~\ref{sec:exp.coverage}, we observe a significant correlation between the invariant structure and fault coverage though this is more dependent on the application rather than the fault type.
However, we find that there is no single class of invariants that provides high coverage across all applications.
This implies that it may be better to derive invariants on an application-specific basis, say based on its algorithm, than to use generic approaches such as \daikon for deriving the invariants.
This is also a direction for future work.

Lastly, we examine the confidence and granularity parameters for the invariant inference module in \techname.
In Section~\ref{sec:confidence}, we observe that lowering the Daikon confidence in \techname does not produce significantly more invariants in the benchmark applications.
This implies that invariants inferred at the default 99\% confidence offers roughly approximate fault coverage to that of invariants at a lower confidence.
In Section~\ref{sec:granularity}, we find that invariants inferred at the basic block granularity do not stabilize within 20 runs, and are, hence, unpreferable for fault detection.
In contrast, the default granularity for \techname is at the function level, which produces stable invariants within only 5 runs.
While changing the granularity to basic blocks may allow more invariants to be inferred overall, the lack of invariant stability within a reasonable number of profiling runs inhibits this option.

\subsection{Improving Fault Coverage}

We suggest two directions to improve \techname's fault coverage.

\textbf{Application Specific Invariants}: 
Unlike the generic classes of invariants inferred by Daikon, used in our inference module, application-specific invariants would be tailored to the program algorithm.
For instance, Daikon currently does not output invariants around the summation of variables.
Such classes of invariants may be important in certain algorithms. Static analysis techniques could be incorporated to determine the appropriate type of invariants for different algorithms. 

\textbf{Changing the Invariant Format}:
While we currently utilize the default invariant  format reported by Daikon, an alternative invariant format may better facilitate fault detection.
For example, invariants are presently reported as mathematical relations between variables.
However, the Order invariant, which has the highest overall fault coverage, reports whether an array is sorted.
Compared to a relational condition between variables, the Order invariant is a stronger condition to satisfy.
 
\subsection{Threats to Validity}

There are three threats to the validity of our results. 
First, \techname uses \daikon for generating likely invariants.
Some results may not apply if an alternate approach to likely invariant generation is used, which is an external threat to validity.
However, as \daikon is the most common likely invariant generator used today, we consider our results valid for most scenarios.

Second, since \techname is limited to tracing local values of primitive data types, the set of generated invariants excludes invariants involving objects and global values.
As a result, the invariants deployed for validation are not necessarily the most relevant invariants for the program.
This is an internal threat to validity.
However, most benchmarks in this study use only primitive data types in their function parameters, and hence this was not an issue in our programs.

Finally, we consider only a limited number of fault types (6) and a limited number of benchmark programs to evaluate \techname.
However, we chose the fault types to cover common software bugs, and hence we believe the results are representative.
Further, we chose the benchmarks to represent a wide variety of scenarios where multithreading is commonly used.


%% file: conclusion.tex
\section{Conclusion}
\label{sec:conclusion}

With processors expanding core counts, multithreaded programs are rising in prevalence. 
Error Propagation Analysis (EPA) is the process of identifying state corruptions and their evolution due to faults. 
Unfortunately, existing methods for EPA that make use of golden traces, are unequipped to handle multithreaded programs due to their inherent non-determinism.

To address this problem, we present an EPA framework that uses likely invariants in lieu of golden traces, to detect faults and track error propagation.
Likely invariants are based on observations of the data values of the program over multiple executions. Our approach, IPA,
instruments programs to automatically derive likely invariants, and then checks for violations of the invariants at runtime, to perform EPA.

We evaluate IPA using 5 benchmark applications. 
Our results indicate that invariants can be dynamically derived in all of our benchmark applications, with reasonable stability, and
 that there is roughly 1 invariant for every 10-100 lines of source code for most applications.
Further, IPA incurs much lower runtime overhead than golden-run based EPA for fault detection, but it incurs a slightly higher setup overhead that
decreases with increasing size of the program.

We inject an assortment of software faults on six different benchmarks to assess the fault coverage offered by the likely invariants.
We find that invariants can capture the effects of faults, but the specific rates of fault coverage differ between different applications and fault types.
In particular, IPA provides high coverage for crashes/hangs across applications, but not so for Silent Data Corruptions (SDCs), which are highly application dependent.
We also find that error propagation varies considerably between single-threaded and multi-threaded versions of the same program, and hence a meaningful comparison
between them is not possible.

Building on these observations, we further dissect the analysis of fault coverage by grouping similar classes of invariants together, 
and find that certain invariant classes offer higher fault coverage than others across different fault classes. However, these differences
are also highly application specific, and there is not a single class of invariant that provides high coverage across all applications.
Further, we explore the choice of \techname's default confidence and granularity parameters in the invariant inference module and find them to be well-justified among alternatives.
Finally, we find that inferring invariants at lower granularities (e.g., basic block) is not able to yield stable invariants, and is hence not viable for fault detection.

In conclusion, likely invariants offer a viable replacement for golden-run based EPA.
The approach performs less well for applications that make substantial use of complex data types, 
which is due to a limitation in the invariant inference engine and could be worked around by 
program transformations that take care of marshalling/unmarshalling of the corresponding data structures when functions operate on them.


%% file: acknowledgements.tex
\section*{Acknowledgements}
\label{sec:acknowledgements}
We sincerely thank the anonymous reviewers for their valuable comments and suggestions.
This work has been partially supported by H2020 CONCORDIA GA \#830927, the Lancaster Security Institute, and by the
Discovery Grants Programme of the Natural Sciences and Engineering Research Council
of Canada (NSERC).


%% file: mainpaper.bbl
\begin{thebibliography}{10}

\bibitem{Duraes2006}
Duraes JA, and Madeira HS.
\newblock {Emulation of Software Faults: A Field Data Study and a Practical Approach}.
\newblock {IEEE} Trans Softw Eng. 2006;{\bf 32}(11):849--867.

\bibitem{Natella2013}
Natella R, Cotroneo D, Duraes JA, and Madeira HS.
\newblock {On Fault Representativeness of Software Fault Injection}.
\newblock {IEEE} Trans Softw Eng. 2013;{\bf 39}(1):80--96.

\bibitem{Aliabadi2014}
Aliabadi MR, Pattabiraman K, and Bidokhti N.
\newblock {Soft-LLFI: A Comprehensive Framework for Software Fault Injection}.
\newblock In: Proc. ISSREW '14; 2014. p. 1--5.

\bibitem{DeMillo1978.coupling}
{DeMillo} RA, {Lipton} RJ, and {Sayward} FG.
\newblock {Hints on Test Data Selection: Help for the Practicing Programmer}.
\newblock Computer. 1978;{\bf 11}(4):34--41.

\bibitem{Offutt1989.coupling}
Offutt A.
\newblock {The Coupling Effect: Fact or Fiction}.
\newblock In: Proceedings of the ACM SIGSOFT '89 Third Symposium on Software Testing, Analysis, and Verification. TAV3. New York, NY, USA: Association for Computing Machinery; 1989. p. 131–140.
\newblock Available from: \url{https://doi.org/10.1145/75308.75324}.

\bibitem{Offutt1992.coupling}
Offutt AJ.
\newblock {Investigations of the Software Testing Coupling Effect}.
\newblock ACM Trans Softw Eng Methodol. 1992 Jan;{\bf 1}(1):5–20.
\newblock Available from: \url{https://doi.org/10.1145/125489.125473}.

\bibitem{Christmansson1998}
Christmansson J, Hiller M, and Rimen M.
\newblock {An experimental comparison of fault and error injection}.
\newblock In: Proc. ISSRE '98; 1998. p. 369--378.

\bibitem{Hiller2002}
Hiller M, Jhumka A, and Suri N.
\newblock {PROPANE: An Environment for Examining the Propagation of Errors in Software}.
\newblock In: Proc. ISSTA '02; 2002. p. 81--85.

\bibitem{Leeke2009}
Leeke M, and Jhumka A.
\newblock {Evaluating the Use of Reference Run Models in Fault Injection Analysis}.
\newblock In: Proc. PRDC '09; 2009. p. 121--124.

\bibitem{Ernst2000}
Ernst MD, Czeisler A, Griswold WG, and Notkin D.
\newblock {Quickly Detecting Relevant Program Invariants}.
\newblock In: Proc. ICSE '00; 2000. p. 449--458.

\bibitem{Schuler2009}
Schuler D, Dallmeier V, and Zeller A.
\newblock {Efficient mutation testing by checking invariant violations}.
\newblock In: Proc. ISSTA '09; 2009. p. 69--80.

\bibitem{Sahoo2013}
Sahoo SK, Criswell J, Geigle C, and Adve V.
\newblock {Using Likely Invariants for Automated Software Fault Localization}.
\newblock SIGPLAN Not. 2013;{\bf 48}(4):139--152.

\bibitem{Perkins2009}
Perkins JH, Kim S, Larsen S, Amarasinghe S, Bachrach J, Carbin M, et~al.
\newblock {Automatically patching errors in deployed software}.
\newblock In: Proc. SOSP '09; 2009. p. 87--102.

\bibitem{Xie2006}
Xie T, and Notkin D.
\newblock {Tool-assisted unit test generation and selection based on operational abstractions}.
\newblock Automated Software Engineering Journal. 2006;{\bf 13}(3):345--371.

\bibitem{Chan2017}
Chan A, Winter S, Saissi H, Pattabiraman K, and Suri N.
\newblock {IPA: Error Propagation Analysis of Multi-Threaded Programs Using Likely Invariants}.
\newblock In: 2017 IEEE International Conference on Software Testing, Verification and Validation (ICST); 2017. p. 184--195.

\bibitem{Koopman2000}
Koopman P, and DeVale J.
\newblock {The exception handling effectiveness of POSIX operating systems}.
\newblock {IEEE} Trans Softw Eng. 2000;{\bf 26}(9):837--848.

\bibitem{Fetzer2004}
Fetzer C, Felber P, and Hogstedt K.
\newblock {Automatic detection and masking of nonatomic exception handling}.
\newblock {IEEE} Trans Softw Eng. 2004;{\bf 30}(8):547--560.

\bibitem{Fu2005}
Fu C, Milanova A, Ryder BG, and Wonnacott DG.
\newblock {Robustness testing of Java server applications}.
\newblock Software Engineering, IEEE Transactions on. 2005;{\bf 31}(4):292--311.

\bibitem{Fu2007}
Fu C, and Ryder BG.
\newblock {Exception-Chain Analysis: Revealing Exception Handling Architecture in Java Server Applications}.
\newblock In: Proc. ICSE '07; 2007. p. 230--239.

\bibitem{Marinescu2009}
Marinescu PD, and Candea G.
\newblock {LFI: A practical and general library-level fault injector}.
\newblock In: Proc. DSN '09; 2009. p. 379--388.

\bibitem{Giuffrida2013}
Giuffrida C, Kuijsten A, and Tanenbaum AS.
\newblock {EDFI: A Dependable Fault Injection Tool for Dependability Benchmarking Experiments}.
\newblock In: Proc. PRDC '13; 2013. p. 31--40.

\bibitem{Sullivan1991}
Sullivan M, and Chillarege R.
\newblock {Software defects and their impact on system availability-a study of field failures in operating systems}.
\newblock In: Proc. FTCS-21; 1991. p. 2--9.

\bibitem{Segall1988}
Segall Z, Vrsalovic D, Siewiorek D, Yaskin D, Kownacki J, Barton J, et~al.
\newblock {FIAT-fault injection based automated testing environment}.
\newblock In: Proc. FTCS-18; 1988. p. 102--107.

\bibitem{LLFI}
Lu Q, Farahani M, Wei J, Thomas A, and Pattabiraman K.
\newblock {LLFI: An Intermediate Code-Level Fault Injection Tool for Hardware Faults}.
\newblock In: Proc. QRS '15; 2015. p. 11--16.

\bibitem{Jin2008}
Jin A.
\newblock {A PIN-Based Dynamic Software Fault Injection System}.
\newblock In: Proc. ICYCS '08. IEEE; 2008. p. 2160--2167.

\bibitem{Lemos}
Lemos GS, and Martins E.
\newblock Specification-guided Golden Run for Analysis of Robustness Testing Results.
\newblock In: Proc. SERE '12; 2012. p. 157--166.

\bibitem{tsan}
{Saissi} H, {Winter} S, {Schwahn} O, {Pattabiraman} K, and {Suri} N.
\newblock {TraceSanitizer - Eliminating the Effects of Non-Determinism on Error Propagation Analysis}.
\newblock In: 2020 50th Annual IEEE/IFIP International Conference on Dependable Systems and Networks (DSN); 2020. p. 52--63.

\bibitem{Gunawi2011}
Gunawi HS, Do T, Joshi P, Alvaro P, Hellerstein JM, Arpaci-Dusseau AC, et~al.
\newblock {FATE and DESTINI: A Framework for Cloud Recovery Testing}.
\newblock In: Proceedings of the 8th USENIX Conference on Networked Systems Design and Implementation. NSDI'11. USA: USENIX Association; 2011. p. 238–252.

\bibitem{Deligiannis2016}
Deligiannis P, McCutchen M, Thomson P, Chen S, Donaldson AF, Erickson J, et~al.
\newblock {Uncovering Bugs in Distributed Storage Systems during Testing (Not in Production!)}.
\newblock In: 14th {USENIX} Conference on File and Storage Technologies ({FAST} 16). Santa Clara, CA: {USENIX} Association; 2016. p. 249--262.
\newblock Available from: \url{https://www.usenix.org/conference/fast16/technical-sessions/presentation/deligiannis}.

\bibitem{Cotroneo2020}
{Cotroneo} D, {De Simone} L, {Liguori} P, and {Natella} R.
\newblock Fault Injection Analytics: A Novel Approach to Discover Failure Modes in Cloud-Computing Systems.
\newblock {IEEE Transactions on Dependable and Secure Computing}. 2020;p. 1--1.

\bibitem{Padon2016}
Padon O, Immerman N, Shoham S, Karbyshev A, and Sagiv M.
\newblock {Decidability of Inferring Inductive Invariants}.
\newblock In: Proc. POPL '16; 2016. p. 217--231.

\bibitem{DIDUCE}
Hangal S, and Lam MS.
\newblock {Tracking Down Software Bugs Using Automatic Anomaly Detection}.
\newblock In: Proc. ICSE '02; 2002. p. 291--301.

\bibitem{DySy}
Csallner C, Tillmann N, and Smaragdakis Y.
\newblock {DySy: Dynamic Symbolic Execution for Invariant Inference}.
\newblock In: Proc. ICSE '08; 2008. p. 281--290.

\bibitem{Kusano2015}
Kusano M, Chattopadhyay A, and Wang C.
\newblock {Dynamic Generation of Likely Invariants for Multithreaded Programs}.
\newblock In: Proc. ICSE '15; 2015. p. 835--846.

\bibitem{Sahoo2008}
Sahoo SK, Li ML, Ramachandran P, Adve SV, Adve VS, and Zhou Y.
\newblock {Using likely program invariants to detect hardware errors}.
\newblock In: Proc. DSN '08; 2008. p. 70--79.

\bibitem{Lu2006}
Lu S, Tucek J, Qin F, and Zhou Y.
\newblock AVIO: Detecting Atomicity Violations via Access Interleaving Invariants.
\newblock In: Proc. ASPLOS XII; 2006. p. 37--48.

\bibitem{mazurkiewicz1987}
Mazurkiewicz A.
\newblock Trace theory.
\newblock In: Petri nets: applications and relationships to other models of concurrency. Springer; 1987. p. 278--324.

\bibitem{Youngs1974}
Youngs EA.
\newblock Human Errors in Programming.
\newblock International Journal of Man-Machine Studies. 1974;{\bf 6}(3):361 -- 376.
\newblock Available from: \url{http://www.sciencedirect.com/science/article/pii/S0020737374800271}.

\bibitem{Ko2003}
Ko AJ, and Myers BA.
\newblock {Development and Evaluation of a Model of Programming Errors}.
\newblock In: Proc. HCC '03; 2003. p. 7--14.

\bibitem{llvm}
Lattner C, and Adve V.
\newblock {LLVM: a compilation framework for lifelong program analysis transformation}.
\newblock In: Proc. CGO '04; 2004. p. 75--86.

\bibitem{PARSEC}
Bienia C.
\newblock Benchmarking Modern Multiprocessors (Dissertation).
\newblock Princeton University; 2011.

\bibitem{nullhttpd}
{Null httpd}.
\newblock Accessed: 2016-05-17;.
\newblock \url{https://sourceforge.net/projects/nullhttpd/.}

\bibitem{nbds}
{Non-blocking data structures}.
\newblock Accessed: 2016-05-17;.
\newblock \url{https://code.go ogle.com/p/nbds/.}

\bibitem{RaiyatAliabadi2016}
Raiyat~Aliabadi M, and Pattabiraman K.
\newblock In: Skavhaug A, Guiochet J, and Bitsch F, editors. FIDL: A Fault Injection Description Language for Compiler-Based SFI Tools. Cham: Springer International Publishing; 2016. p. 12--23.
\newblock Available from: \url{http://dx.doi.org/10.1007/978-3-319-45477-1_2}.

\bibitem{V2005}
Vipindeep V, and Jalote P. List of Common Bugs and Programming Practices to avoid them; 2005.

\bibitem{Hsueh1997}
Hsueh MC, Tsai TK, and Iyer RK.
\newblock Fault Injection Techniques and Tools.
\newblock {IEEE} Computer. 1997;{\bf 30}(4):75--82.
\newblock Available from: \url{http://dx.doi.org/10.1109/2.585157}.

\bibitem{Ghosh1998}
Ghosh AK, O'Connor T, and McGraw G.
\newblock An automated approach for identifying potential vulnerabilities in software.
\newblock In: Proc. IEEE S\,\&\,P; 1998. p. 104--114.

\bibitem{Jeffrey2008}
Jeffrey D, Gupta N, and Gupta R.
\newblock Identifying the root causes of memory bugs using corrupted memory location suppression.
\newblock In: Proc. ICSM '08; 2008. p. 356--365.

\bibitem{Spinellis2016}
Spinellis D, Louridas P, and Kechagia M.
\newblock The Evolution of C Programming Practices: A Study of the Unix Operating System 1973--2015.
\newblock In: Proceedings of the 38th International Conference on Software Engineering. ICSE '16. New York, NY, USA: Association for Computing Machinery; 2016. p. 748–759.
\newblock Available from: \url{https://doi.org/10.1145/2884781.2884799}.

\end{thebibliography}
